\documentclass[10pt]{iopart}
\pdfoutput=1
\usepackage{iopams,amsfonts,amssymb,amsthm}
\usepackage{rawfonts}
\usepackage[usenames]{color}
\usepackage{subfigure}

\input prepictex
\input pictex
\input postpictex

\def\C#1{{\cal #1}}
\def\Pr{\hbox{Pr}}
\def\mean#1{{\left\langle #1 \right\rangle}}

\def\SMALL{\scriptsize}

\def\arcsec{{\,\hbox{arcsec}\,}}

\newcommand{\NatN}{{\mathbb N}}
\newcommand{\RealN}{{\mathbb R}}

\newcommand{\ELL}{{\mathbb L}}

\newcommand{\LH}{\left[} \newcommand{\RH}{\right]}
\newcommand{\LV}{\left|} \newcommand{\RV}{\right|}
\newcommand{\LB}{\left(} \newcommand{\RB}{\right)}
\newcommand{\LL}{\left\langle} \newcommand{\RR}{\right\rangle}
\newcommand{\LC}{\left\{} \newcommand{\RC}{\right\}}

\newcommand{\ol}[1]{\overline{#1}}

\newcommand{\Ref}[1]{(\ref{#1})}

\usepackage{multirow}
\usepackage{graphicx}

\begin{document}

\title[Minimal Knots]{Minimal Knotted Polygons in
Cubic Lattices}

\author{E J Janse van Rensburg$\dagger$\footnote[3]{To whom 
correspondence should be addressed (\texttt{rensburg@yorku.ca)}}
and A Rechnitzer$\ddagger$}

\address{$\dagger$Department of Mathematics and Statistics, 
York University\\ Toronto, Ontario M3J~1P3, Canada\\
\texttt{rensburg@yorku.ca}}

\address{$\ddagger$Department of Mathematics, 
The University of British Columbia\\
Vancouver V6T~1Z2, British Columbia , Canada\\
\texttt{andrewr@math.ubc.ca}}

\begin{abstract}
An implementation of BFACF-style algorithms \cite{AC83,ACF83,BF81} 
on knotted polygons in the simple cubic, face centered cubic and
body centered cubic lattice \cite{JvRR10,JvRR11} is used to estimate 
the statistics and writhe of minimal length knotted polygons in 
each of the lattices.  Data are collected and analysed on 
minimal length knotted polygons, their entropy, and their 
lattice curvature and writhe.
\end{abstract}

\maketitle

%%%%%%%%%%%%%%%%%%%%%%%%%%%%%%%%%%%%%%%%%%%%%%%%%%%%%%%%%%%%%%%%
%%%%%%%%%%%%%%%%%%%%%%%%%%%%%%%%%%%%%%%%%%%%%%%%%%%%%%%%%%%%%%%%
\section{Introduction}

A lattice polygon is a model of ring polymer in a good solvent,
and is a useful in the examination of the entropy properties of
ring polymers in dilute solution \cite{F69,deG79}.  Ring polymers
can be knotted \cite{FW61,D62,MW86}, and this topological property
can be modeled by examining knotted polygons in a three dimensional
lattice.  The effects of knotting (and linking) on the entropic 
properties of knotted ring polymers remains little understood, apart
from empirical data collected via experimentation on knotted
ring polymers (for example, knotted DNA molecules \cite{TAVSR01})
or by numerical simulations of models of knotted ring polymers
\cite{P89,BOS07}).  Knots in polymers are generally thought to
have effects on both the physical \cite{UKAYK07} and thermodynamic
properties \cite{V95} of the polymer, but these effects are 
difficult to understand in part because the different knot types 
in the polymer may have different properties.

\begin{figure}[h!]
\centering
\parbox{1.2in}{%
 \includegraphics[scale = 0.1] {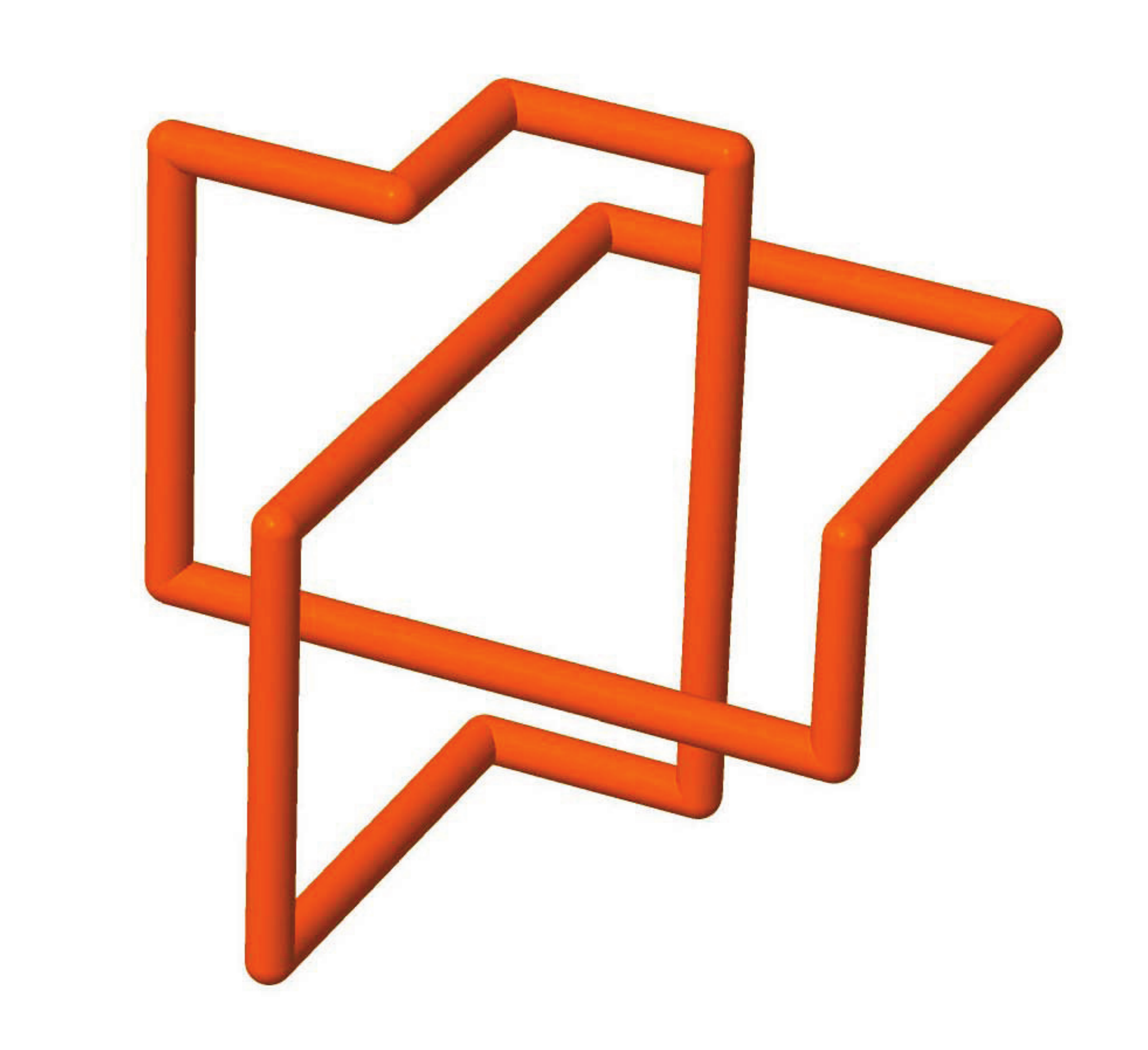}%
%  \includegraphics[scale = 0.1] {Jmol31.eps}%
   %\caption{One}%
   \label{fig:subfig1}}%
\hspace{11mm}%
\begin{minipage}{1.2in}%
  \includegraphics[scale = 0.11] {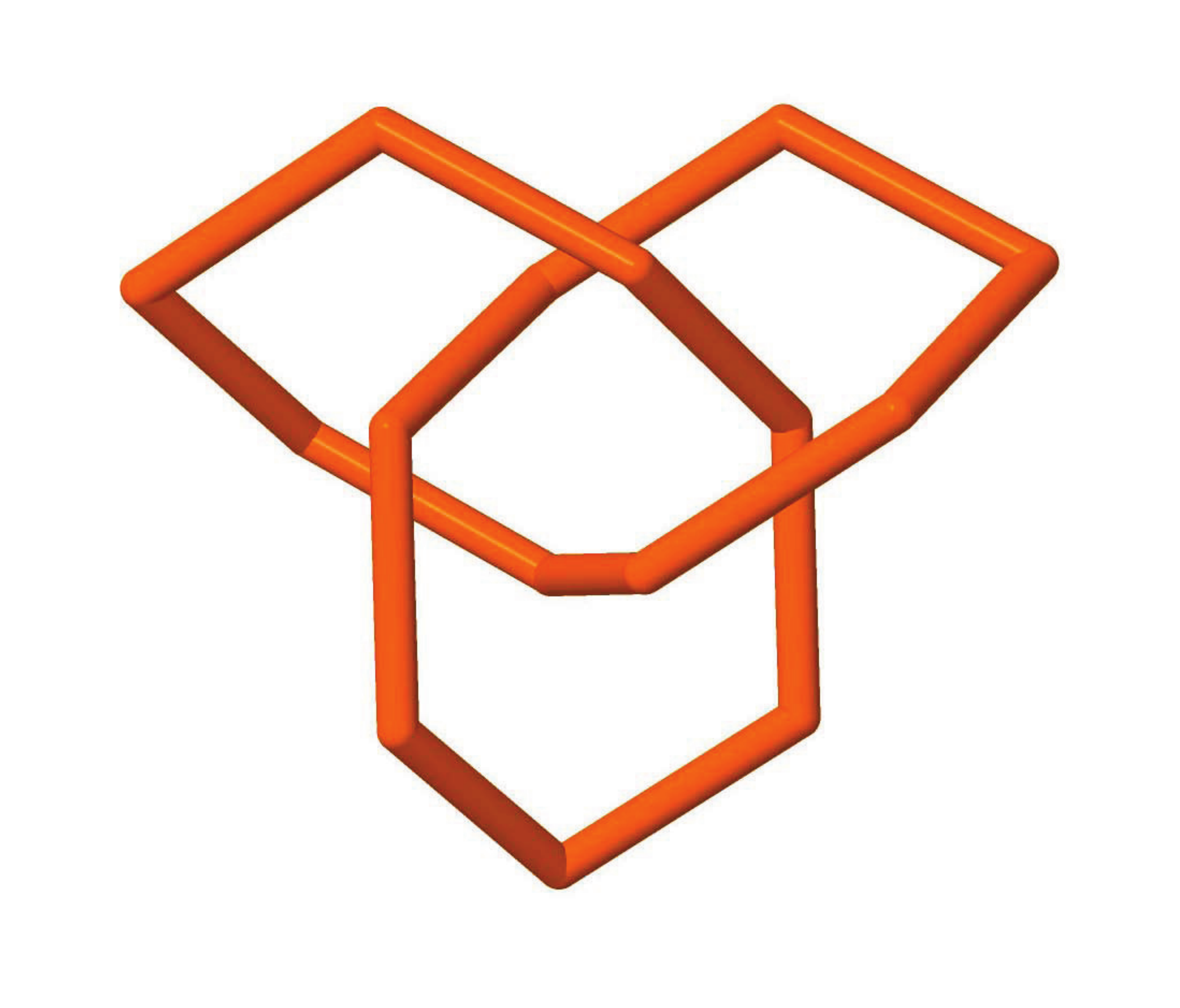}%
%   \includegraphics[scale = 0.11] {Jmolfcc31.eps}%
    %\caption{Second}%
    \label{fig:subfig2}%
\end{minipage}%
\hspace{12mm}%
\begin{minipage}{1.2in}%
  \includegraphics[scale = 0.11] {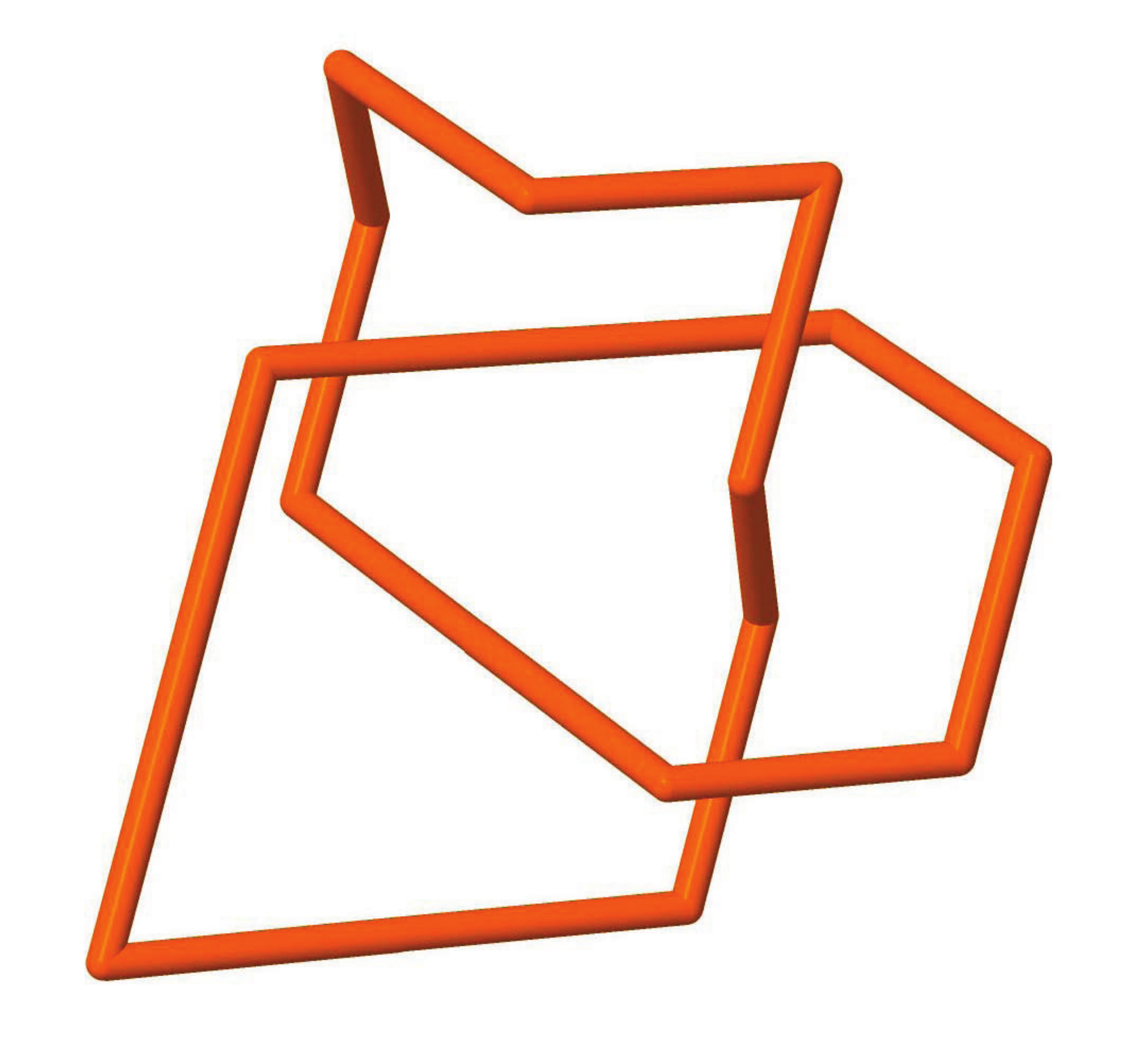}%
    %\caption{Second}%
    \label{fig:subfig3}%
\end{minipage}%
\caption{Lattice trefoils (of knot type $3_1^+$ in the standard knot
tables \cite{BZ85}) in three dimensional cubic lattices.  On the
left a lattice trefoil is embedded in the simple cubic lattice.
In the middle a lattice trefoil in the face-centered cubic
lattice is illustrated, while the right panel is a realisation
of a lattice trefoil in the body-centered cubic lattice.}
\label{FIG1} %%ZXZ[FIG1]
\end{figure}

A polygon in a regular lattice $\ELL$ is composed of a sequence
of distinct vertices $\{a_0,a_1,\ldots,a_n\}$ such that
$a_ja_{j+1}$ and $a_na_0$ are lattice edges for each
$j=0,1,\ldots,n-1$.  Two polygons are said to be equivalent
if the first is translationally equivalent to the second.  
Such equivalence classes of polygons are unrooted, and we abuse 
this terminology by referring to these equivalence classes 
as {\it (lattice) polygons}. In figure \ref{FIG1} we display three
polygons in regular cubic lattices.  The polygon on the left is a
lattice trefoil knot in the simple cubic (SC) lattice.  In the
middle a lattice trefoil is displayed in the face-centered cubic
(FCC) lattice, while in the right hand panel an example of a lattice
trefoil in the body-centered cubic (BCC) lattice is illustrated.

A lattice polygon has length $n$ if it is composed of $n$ edges
and $n$ vertices. The number of lattice polygons of length $n$ 
is the number of distinct polygons of length $n$, denoted by 
$p_n$.  The function $p_n$ is the most basic combinatorial 
quantity associated with lattice polygons, and $\log p_n$ is
a measure of the entropy of the lattice polygon at length $n$. 

Determining $p_n$ in regular lattices is an old and difficult
combinatorial problem \cite{H61}. Observe that $p_{2n+1}=0$ for 
$n\in\NatN$ in the SC lattice, and it known that the growth 
constant $\mu$ defined by the limit
\begin{equation}
\lim_{n\to\infty} p_n^{1/n} = \mu > 0
\label{eqn1AA} %%ZXZ{eqn1AA}
\end{equation}
exists and is finite in the SC lattice \cite{H61} if the limit 
is taken through even values of $n$. This result can be extended 
to other lattices, including the FCC and the BCC
lattices, using the same basic approach in reference \cite{H61}
(and by taking limits through even numbers in the BCC). 
In the hexagonal lattice it is known that 
$\mu = \sqrt{2+\sqrt{2}}$ \cite{DCS10}.   

In three dimensional lattices polygons are models of ring polymers.
Knotted polygons are similarly a model of knotted ring polymers,
see for example reference \cite{FW61} on the importance of
topology in the chemistry of ring polymers, and \cite{D62} 
on the occurrence of knotted conformations in DNA.
  
%%%%%%%%%%%%%%%%%%%%%%%%%%%%%
%%%%%%%%%%%%%%%%%%%%%%%%%%%%%
\subsection{Knotted Polygons}

Let $S^1$ be a circle.  An embedding of $S^1$ 
into $\RealN^3$ is an injection $f: S^1 \to \RealN^3$.  We say 
that $f$ is tame if it contains no singular points, and a 
tame embedding is piecewise linear and finite if the image 
of $f$ is the union of line segments of finite length in $\RealN^3$.  
A tame piecewise linear embedding of $S^1$ into $\RealN^3$ 
is is also called a {\it polygon}.

Tame embeddings $S^1$ into $\RealN^3$ are tame knots, 
and the set of polygons compose a class of piecewise 
linear knots denoted by $\C{K}_p$.  If the class of all
lattice polygons (for example, in a lattice $\ELL$) is
denoted by $\C{P}$, then $\C{P} \subset \C{K}_p$ so that
each lattice polygon is also a tame and piecewise linear
knot in $\RealN^3$.  This defines the knot type $K$
of every polygon in a unique way.  In particular, two
polygons in $\C{P}$ are equivalent as knots if and only
if they are ambient isotopic as tame knots in $\C{K}_p$.

Define $p_n (K)$ to be the number of lattice polygons in
$\ELL$, of length $n$ and knot type $K$, counted modulo
equivalence under translations in $\ELL$.  Then $p_n(K)$
is the number of unrooted lattice polygons of length $n$ 
and knot type $K$. Observe that $p_n (K) = 0$ if $n$ is odd,
and hence, consider $p_n(K)$ to be a function on even numbers;
$p_n(K):\,2\NatN \to \NatN$.

It follows that $p_n(0_1) = 0$ if $n<4$ and $p_4(0_1)=3$ in 
the SC lattice where $0_1$ is the unknot (the simplest knot
type). If $K\not=0_1$ is not the unknot, then
in the SC lattice it is known that $p_n(K)=0$ if $n<24$ and 
that $p_{24}(K)>0$ \cite{D93}.  In particular, $p_{24}(3_1)
=3328$ \cite{PDSAV10} while $p_n(K)=0$ if $K\not= 0_1$
or $K\not= 3_1$.  

It is known that
\begin{equation}
\limsup_{n\to\infty} \LH p_n(K)\RH^{1/n}
= \ol{\mu}_K < \mu
\label{eqn2} %%ZXZ[eqn2]
\end{equation}
in the SC lattice; see reference \cite{SuW88}. If $K=0_1$
is the unknot, then it is known that
\begin{equation}
\lim_{n\to\infty} \LH p_n(0_1)\RH^{1/n}
= \mu_0 < \mu 
\label{eqn3} %%ZXZ[eqn3]
\end{equation}
and it follows in addition that $\mu_0 \leq \ol{\mu}_K < \mu$; see 
for example \cite{JvR02,JvR08}.  There are substantial numerical 
evidence in the literature that $\mu_0 = \ol{\mu}_K$ for all knot 
types $K$ (see reference \cite{JvR08} for a review, and references
\cite{OJ96,JvRR08,JvRR11a} for more on this). Overall, these results 
are strong evidence that the asymptotic behaviour of $p_n(K)$ is 
given to leading order to
\begin{equation}
p_n (K) \simeq C_K n^{\alpha_0 + N_K - 3} \mu_0^n,
\label{eqn4} %%ZXZ[eqn4]
\end{equation}
where $N_K$ is the number of prime components of knot type
$K$, and $\alpha_0$ is the entropic exponent which is independent
of knot type.  The amplitude $C_K$ is dependent on the knot
type $K$.  In particular, simulations show that the amplitude
ratio $\LH p_n(K)/p_n(L)\RH \to \LH C_K/C_L \RH \not= 0$ if
$N_K = N_L$ \cite{OJ96}; this strongly supports the proposed
scaling in equation \Ref{eqn4}.

\begin{figure}[t]
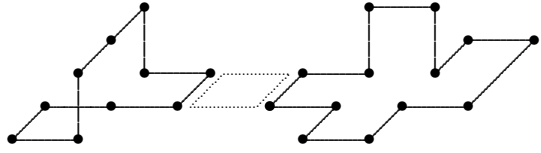

\input fig1.inp
\caption{Concatenating polygons in the SC lattice.  The 
{\it top edge} of the polygon on the left is defined at that
edge with lexicographic most midpoint, and the {\it bottom
edge} of the polygon on the right as that edge with lexicographic
least midpoint.  By translating, and rotating the polygon on 
the right until its bottom edge is parallel to the top edge
of the polygon on the left, and translated one step in
the $X$-direction, the two polygons can be concatenated
into a single polygon by inserting the dotted polygon of
length four between the two, as illustrated, and then 
deleting edges which are doubled up. If the polygon
on the left has length $n$ and knot type $K$, then it can 
be chosen in $p_n (K)$ ways, and if the polygon on the
right has length $m$ and knot type $L$, then it can be
chosen in $p_m(L)/2$ ways, since its bottom edge much be
parallel to the top edge of the polygon on the left.
This shows that $p_n(K) p_m(L) \leq 2\, p_{n+m}(K\# L)$,
since the concatenated polygon has length $n+m$ and
knot type the connected sum $K\# L$ of the the knot types
$K$ and $L$. This construction generalises in the obvious
way to the FCC and BCC lattices.}
\label{fig1} %%ZXZ[fig1]
\end{figure}

Growth constants for knotted polygons in the FCC and BCC 
in equations \Ref{eqn2} and \Ref{eqn3} have not been examined
in the literature, but there is general agreement that the
methods of proof in the SC lattice will demonstrate these
same relations in the FCC and BCC.  In particular, by
concatenating SC lattice polygons as schematically illustrated in 
figure \ref{fig1}, it follows that 
\begin{equation}
p_n (K) p_m (L) \leq 2\, p_{n+m} (K\# L)
\end{equation}
where $K\# L$ is the connected sum of the knot types $K$ and $L$.

%%%%%%%%%%%%%%%%%
%%%%%%%%%%%%%%%%

Similar results are known in the FCC lattice:  One has that
$p_n(0_1)=0$ if $n<3$, and $p_3(0_1) = 8$.  Similarly,
$p_n(3_1)=0$ if $n<15$, while $p_{15}(3_1) = 64$.  Observe that
in the FCC lattice, $p_n (K)$ is a function on $\NatN$; $p_n(K):\,
\NatN\to\NatN$.  That is, there are polygons of odd length.

The construction in figure \ref{fig1} generalises to the
FCC lattice. In this case, the {\it top vertex} of the polygon
is that vertex with lexicographic most coordinates.  The top 
vertex $t$ is incident with two edges, and the top edge is that
edge with midpoint with lexicographic most coordinates. 
The top edge of a FCC polygon is parallel to one of six possible 
directions, giving six different classes of polygons.  One of
these classes is the most numerous, containing at least
$p_n(K)/6$ polygons and with top edge parallel to (say)
direction $\beta$, if the polygons has length $n$ and knot 
type $K$.

Similarly, the bottom vertex and bottom edge of a FCC polygon
of length $m$ and knot type $L$ can be identified, and there
is a direction $\gamma$ such that the class of FCC polygons
with bottom edge is parallel to $\gamma$ is the most numerous
and is at least $p_m (L)/6$.

By choosing a polygon of knot type $K$, top vertex $t$ and 
length $n$ with top edge parallel to $\beta$, and a second 
polygon of length $m$, bottom vertex $b$, with bottom edge 
parallel to $\gamma$, these polygons can be concatenated
similarly to the construction in figure \ref{fig1} by inserting
a polygon of length (say) $N+2$ between them. Accounting
for the number of choices of the polygons on the left and right,
and for the change in the number of edges, this shows that 
\begin{equation}
p_n (K) p_m (L) \leq 36\, p_{n+m+N} (K \# L)
\label{eqn6} %%ZXZ[eqn6]
\end{equation}
in the FCC, where $N$ is independent of $n$ and $m$ .  The
polygon of length $N+2$ is inserted to join the top and bottom
edges of the respective polygons, since they may not be parallel
a priori to the concatenation.  Some reflection shows that
the choice $N=2$ is sufficient in this case.

The relation in equation \Ref{eqn6} shows that $[p_{n-N}/36]$ 
and $[p_{n-N} (0_1)/36]$ are supermultiplicative functions 
in the FCC, and this proves existence of the limits
$\lim_{n\to\infty} \LH p_n \RH^{1/n} = \mu$ and 
$\lim_{n\to\infty} \LH p_n (0_1)\RH^{1/n} = \mu_0$
in the FCC \cite{H48}.  In addition, with $\ol{\mu}_K$ defined in 
the FCC as in equation \Ref{eqn2}, it also follows from
equation \Ref{eqn6} that $\mu_{0_1} \leq \ol{\mu}_K \leq \mu$.
That $\ol{\mu}_K < \mu$ would follow from a pattern theorem for
polygons in the FCC (and it is widely expected that the methods
in reference \cite{K63,K64} will prove a pattern theorem for
polygons in the FCC).

In the BCC lattice one may verify that $p_n(0_1)=0$ if $n<4$, 
and $p_4(0_1) = 12$.  Similarly, $p_n(3_1)=0$ if $n<18$, 
while $p_{18}(3_1) = 1584$.  Observe that in the BCC lattice 
$p_n (K)$ is a function on even numbers; $p_n(K):\,
2\NatN\to\NatN$, similar to the case in the SC lattice.

Finally, arguments similar to the above show that in the BCC lattice
there exists an $N$ independent of $n$ and $m$ such that
\begin{equation}
p_n(K) p_m (L) \leq 16\, p_{n+m+N} (K\# L) .
\label{eqn7} %%ZXZ[eqn7]
\end{equation}
Thus, in the BCC one similarly expects that 
$\lim_{n\to\infty} \LH p_n \RH^{1/n} = \mu$ and 
$\lim_{n\to\infty} \LH p_n (0_1)\RH^{1/n} = \mu_0$
exists in the BCC, and with $\ol{\mu}_K$ defined in the BCC 
as in equation \Ref{eqn2}, it also follows from
equation \Ref{eqn6} that $\mu_0 \leq \ol{\mu}_K \leq \mu$.
Similarly, a pattern theorem will show that $\ol{\mu}_K < \mu$.
In the BCC one may choose $N=2$.

Generally, these results are consistent with the hypothesis
that $\ol{\mu}_K = \mu_0$ in the BCC and FCC lattices, while the 
asymptotic form for $p_n (K)$ in equation \Ref{eqn4} is 
expected to apply in these lattices as well.  By computing
amplitude ratios $\LH C_K / C_L \RH$ in reference
\cite{JvRR11} for a selection of knots, strong numerical evidence
for equation \Ref{eqn4} in the BCC and FCC were obtained.

%%%%%%%%%%%%%%%%%%%%%%%%%%%%%%%%%%%%%%%%%%%%%%%%%%%%%%%%%%%%
%%%%%%%%%%%%%%%%%%%%%%%%%%%%%%%%%%%%%%%%%%%%%%%%%%%%%%%%%%%%
\subsection{Minimal Length Knots and the Lattice Edge Index}

Given a knot type $K$ there exists an $n_K$ such that
$p_{n_K} (K) >0$ but $p_n(K) = 0$ for all $n< n_K$.
The number $n_K$ is the minimal length of the knot type $K$ in
the lattice \cite{D93,JvRP95}.  For example, if 
$K = 3_1^+$ (a right-handed trefoil knot) then in the
SC lattice it is known that $p_{24} (3_1^+) = 1664$
while $p_n (3_1^+) = 0$ for all $n < 24$.  Thus $n_{3_1^+}=24$ is 
the minimal length of (right-handed) trefoils in the SC
lattice \cite{DIASV09}.  Observe that $n_{3_1^+}=n_{3_1^-}
\,(=n_{3_1})$, and this is generally true for all knot types.

Similar results are not available in the BCC and FCC, although
numerical simulations have shown that $n_{3_1^+} = 18$ in the
BCC and $n_{3_1^+} = 15$ in the FCC \cite{JvRR10,JvRR11,JvRR11a}.

The construction in figure \ref{fig1} shows that
\begin{equation}
n_{K \# K} \leq 2\, n_{K} \,\,\hbox{and}\,\,
n_{K \# L} \leq  n_{K} + n_{L}
\end{equation}
in the SC lattice.  More generally, observe that for non-negative
integers $p$, 
\begin{equation}
n_{K^p} \leq p\, n_K .
\end{equation}
This in particular shows that the {\it minimal lattice
edge index} defined by
\begin{equation}
\inf_p \LH \frac{n_{K^p}}{p} \RH = 
\lim_{p\to\infty} \LH \frac{n_{K^p}}{p} \RH = \alpha_K
\label{eqn10} %%ZXZ[eqn10]
\end{equation}
exists, and moreover, $\alpha_K \geq 4(b_K -1)$, where
$b_K$ is the bridge number of the knot type $K$; see references
\cite{JvRP95,JvR96,JvR08} for details. Since $b_K \geq 2$ if
$K \not= \emptyset$, it follows that $\alpha_K \geq 4$ for
non-trivial knots types in the SC lattice. Observe
that $\alpha_{0_1} = 0$ and that it is known that
$4 \leq \alpha_{3_1^+} \leq 17$ \cite{JvRP95,JvR96}. 

In the BCC and FCC lattices one may consult equations \Ref{eqn6} and
\Ref{eqn7} to see that for non-negative integers, 
\begin{equation}
n_{K^p \# K^q} \leq n_{K^p} + n_{K^q} + N.
\end{equation}
Thus, $n_{K^p}+N$ is a subadditive function of $p$, and hence
\begin{equation}
\inf_p \LH \frac{n_{K^p}+N}{p} \RH =
\lim_{p\to\infty} \LH \frac{n_{K^p}}{p} \RH = \alpha_K
\end{equation}
exists \cite{H48}.  Moreover, as in the SC lattice, one may 
present arguments similar to those in the proof of theorem 2
in reference \cite{JvR99} to see that $\alpha_K \geq 3(b_K -1)$
in the FCC and $\alpha_K \geq 2(b_K-1)$ in the BCC.  Hence, if 
$K$ is not the unknot, then $\alpha_K \geq 3$ in the FCC 
and $\alpha_K \geq 2$ in the BCC.

We shall also work with the total number of distinct knot types
$K$ with $n_K \leq n$, denoted by $Q_n$.  It is known that
$Q_n = 1$ if $n < 24$ in the SC lattice, and that $Q_n = 3$ 
if $24 \leq n < 30$ \cite{DIASV09}, also in the SC lattice.
$Q_n$ grows exponentially with $n$.

%%%%%%%%%%%%%%%%%%%%%%%%%%%%%%%%%%%%%%%%%%%%%%%%%%%%%%%%%%%
%%%%%%%%%%%%%%%%%%%%%%%%%%%%%%%%%%%%%%%%%%%%%%%%%%%%%%%%%%%
\subsection{The Entropy of Minimal Length Knotted Polygons}

If $n=n_K$, then $p_n(K) > 0$ for a given knot type.  The
{\it entropy} of the knot type $K$ at minimal length is
given by $\log p_n (K)$ when $n=n_K$.\footnote{Sometimes,
this notion will be abused when we refer to $p_n(K)$ as the 
(lattice) entropy of polygons of length $n$ and knot type $K$.}  
More generally, the entropy of lattice knots of minimal length
and knot type $K$ can be studied by defining the {\it density} of the 
knot type $K$ at minimal length by
\begin{equation}
\C{P}_K = p_{n_K} (K) .
\end{equation}
Then one may verify that $\C{P}_\emptyset = 3$ in the SC lattice,
and $\C{P}_\emptyset = 12$ in the BCC lattice while
$\C{P}_\emptyset = 8$ in the FCC lattice. 

It is also known that $\C{P}_{3_1^+} = 1664$ in the SC lattice
\cite{DIASV09}. Since $3_1$ is a chiral knot type, it follows that 
the total number of minimal length lattice polygons of knot type 
$3_1$ is given by $\C{P}_{3_1} = \C{P}_{3_1^+} + \C{P}_{3_1^-} = 3328$.

Generally there does not appear to exist simple relations between
$\C{P}_K$ and $\C{P}_{K^m}$.  However, $\C{P}_{K^m}$ should
increase exponentially with $m$, since $n_{K^m}$ is bounded
linearly with $m$ if $K$ is a non-trivial knot type \cite{JvR99}.  
Thus, the {\it entropic index per knot component of 
the knot type $K$} can be  defined by
\begin{equation}
\limsup_{m\to\infty} \LH \frac{\log \C{P}_{K^m}}{m} \RH = \gamma_K .
\label{eqn14} %%ZXZ[eqn14]
\end{equation}
Obviously, since $\C{P}_{\emptyset^m} = 4$ for all values
of $m$, it follows that $\gamma_\emptyset = 0$.  Also,
$\gamma_K \geq 0$ for all knot types $K$.  Showing that 
$\gamma_K > 0$ for all non-trivial knot types $K$ is an 
open question.

The collection of $\C{P}_K$ minimal length lattice knots are 
partitioned in symmetry (or equivalence) classes by rotations 
and reflections (which compose the octahedral group, which is 
the symmetry group of the cubic lattices).  Since the group has 
$24$ elements, each symmetry class may contain at most $24$ 
equivalent polygons.  The total number of symmetry classes of
minimal length lattice knots of type $K$ is denoted by $\C{S}_K$.
For example, in the SC lattice it is known that $\C{S}_{0_1} = 1$ 
and this class has $3$ minimal length lattice knots of length $4$.
It has been shown that $\C{S}_{3_1} = 142$ in the SC, of which 
$137$ classes have $24$ members each and $5$ have $8$ members 
each \cite{DIASV09}.

%%%%%%%%%%%%%%%%%%%%%%%%%%%%%%%%%%%%%%%%%%%%%%%%%%%%%%%%%%%%%%
%%%%%%%%%%%%%%%%%%%%%%%%%%%%%%%%%%%%%%%%%%%%%%%%%%%%%%%%%%%%%%
\subsection{The Mean Absolute Writhe of Minimal Length Knotted Polygons}

The writhe of a closed curve is a geometric measure of its
self-entanglement. It is defined as follows:  The projection
of a closed curve in $\RealN^3$ onto a geometric plane is
{\it regular} if all multiple points in the projection are
double points, and if projected arcs intersect transversely
at each double point.  

\begin{figure}[t]
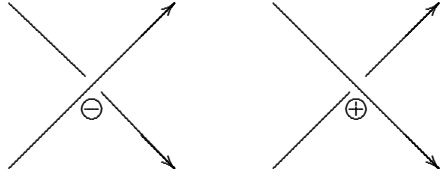

\input fig2.inp
\caption{A negative crossing (left) and a positive crossing
(right) of the intersections in a regular projections of
a simple closed curved.  The signs are assigned using a
right hand rule.}
\label{fig2} %%ZXZ[fig2]
\end{figure}

Intersections (referred to as ``crossings") in a regular
projection are signed by the use of a right hand rule:
The curve is oriented and the sign is assigned as illustrated
in figure \ref{fig2}.  The writhe of the projected curve
is the sum of the signed crossings.  The writhe of the space
curve is the average writhe over all possible regular
projections of the curve.  For a lattice polygon $\omega$ this is
defined by
\begin{equation}
W_r (\omega) = \frac{1}{4\pi} \int_{u\in S^2} W_r (\omega,u)
\end{equation}
where $W_r (\omega,u)$ is the writhe of the projection along
the unit vector $u$ (which takes values in the unit sphere
$S^2$ -- this is called the \textit{writhing number} of the 
projection).  This follows because almost all projections of
$\omega$ are regular.

%%%%%%%%%%%%%%%%%
%%%%%%%%%%%%%%%%

The writhe of a closed curve was introduced by Fuller \cite{F71}.
It was shown by Lacher and Sumners \cite{LS91} that the writhe
of a lattice curve is given by the average of the linking 
number of $\omega$ with its push-offs $\omega+\epsilon u$, for 
$u\in S^2$, and $\epsilon>0$ small. That is,
\begin{equation}
W_r (\omega) = \frac{1}{4\pi} \int_{u\in S^2}
L_k (\omega,\omega + \epsilon u).
\end{equation}
In the SC lattice, this simplifies to the average linking number
of $\omega$ with four of its push-offs into non-antipodal octants:
\begin{equation}
W_r (\omega) = \frac{1}{4} \sum_{i=1}^4 L_k (\omega,\omega+u_i)
\end{equation}
where, for example, one may take $u_1 = (0.5,0.5,0.5)$,
$u_2 = (0.5,-0.5,0.5)$, $u_3 = (-0.5,0.5,0.5)$ and
$u_4 = (-0.5,-0.5,0.5)$.  This shows that $4\,W_r(K)$ is
an integer.

The average writhe $\langle W_r (K) \rangle_n$ of polygons
of knot type $K$ and length $n$ is defined by
\begin{equation}
\langle W_r (K) \rangle_n = \frac{1}{p_n(K)}
\sum_{|\omega|=n} W_r (\omega)
\end{equation}
where the sum is over all polygons of length $n$ and knot type
$K$.  If $K$ is an achiral knot, then 
$\langle W_r (K) \rangle_n =0$ for each value of $n$ \cite{JvROSTW96}.

The average absolute writhe $\langle \LV W_r (K)\RV \rangle_n$
of polygons of knot type $K$ and length $n$ is defined by
\begin{equation}
\langle\LV W_r (K)\RV \rangle_n = \frac{1}{p_n(K)}
\sum_{|\omega|=n} \LV W_r (\omega) \RV
\end{equation}
where the sum is over all polygons of length $n$ and knot type $K$.  

The averaged writhe $\C{W}_K$ and the average absolute writhe 
$|\C{W}|_K$ of lattice knots of both minimal length and knot 
type $K$ are defined as the average and average absolute writhe 
of polygons of knot type $K$ and minimal length:
\begin{equation}
\C{W}_K = \langle W_r (K) \rangle_n {\hbox{\Large$|$}}_{n=n_K}; \quad
\C{|W|}_K = \langle\LV W_r (K)\RV \rangle_n {\hbox{\Large$|$}}_{n=n_K}.
\end{equation}

The writhe of polygons in the BCC and FCC lattices can also be
determined by computing linking numbers between polygons
and their push-offs \cite{LS06}.  Normally, the writhes
in these lattices are related to the average writhing numbers
of projections of the polygons onto planes normal to a set
of given vectors. 

The writhe of a polygon in the FCC lattice is normally an irrational
number \cite{GIW99}.  The prescription for determining the writhe 
of polygons in the FCC lattice can be found in reference \cite{LS06}
and is as follows:  Put $\alpha = 3 \arcsec 3 - \pi$ and
$\beta = (\pi/2-\alpha)/3$.  Then the writhe of a polygon
$\omega$ is given by
\begin{equation}
W_r (\omega) = \frac{1}{2\pi}
\LB \alpha \sum_{i=1}^4 W_r (\omega,u_i)
+ \beta \sum_{i=1}^8 W_r (\omega,v_i) \RB
\label{eqn21} %%ZXZ[eqn21]
\end{equation}
where the vectors $u_i$ are defined by 
$u_i = (\pm 3/\sqrt{22},\pm 3/\sqrt{22},2/\sqrt{22})$ for all 
possible choices of the signs, and the vectors $v_i$ are defined by
$v_i = (\pm \sqrt{5}/\sqrt{6},\pm 1/\sqrt{30},2/\sqrt{30})$,
$(\pm 1/\sqrt{30},\pm \sqrt{5}/\sqrt{6},2/\sqrt{30})$,
$(\pm 1/\sqrt{38},\pm 1/\sqrt{38},6/\sqrt{38})$ again
for all possible choices of the signs. The writhing number
$W_r (\omega,u_i)$ of $\omega$ is defined as before as
the sum of the signed crossings in the projected $\omega$ on a 
plane normal to $u_i$.  

In the BCC lattice the writhe of a polygon $\omega$ can be
computed by
\begin{equation}
W_r (\omega) = \frac{1}{12} \sum_{i=1}^{12} W_r (\omega,u_i)
\label{eqn21AA} %%ZXZ[eqn21AA]
\end{equation}
where the vectors $u_i$ are defined by
$u_i = (\pm 1/\sqrt{10},3/\sqrt{10},0)$,
$(\pm 1/\sqrt{10},0,3/\sqrt{10})$,
$(0,\pm 1/\sqrt{10},3/\sqrt{10})$,
$(0,3/\sqrt{10},\pm 1/\sqrt{10})$,
$(3/\sqrt{10},\pm 1/\sqrt{10},0)$,
$(3/\sqrt{10},0,\pm 1/\sqrt{10})$, for all possible choices
of the signs.

By appealing to the Calugareanu and White formula $L_k = T_w + W_r$
\cite{C61,W69} for a ribbon, one can compute $W_r (\omega,u_i)$
by creating a ribbon $(\omega,\omega+\epsilon u_i)$ with boundaries 
$\omega$ and $\omega+\epsilon u_i$ (this is a \emph{push-off} of 
$\omega$ by $\epsilon$ in the (constant) direction of $u_i$).  
Since the twist of this ribbon is zero, one has that 
$W_r (\omega,u_i) = W_r (\omega+\epsilon u_i,u_i) 
= L_k (\omega,\omega+\epsilon u_i)$, 
and the writhe can be computed by the linking number of the 
knot $W_r (\omega,u_i)$ and its push-off 
$W_r (\omega,u_i)+\epsilon u_i$.

Equation \Ref{eqn21AA} shows that $12\,W_r(\omega)$ is an integer
in the BCC lattice.  Thus, the mean writhe of a finite collections
of polygons in the BCC lattice is a rational number.

%%%%%%%%%%%%%%%%%%%%%%%%%%%%%%%%%%%%%%%%%%%%%%%%%%%%%%%%%%%%%%%%%%
%%%%%%%%%%%%%%%%%%%%%%%%%%%%%%%%%%%%%%%%%%%%%%%%%%%%%%%%%%%%%%%%%%
\subsection{Curvature of Lattice Knots}

The total curvature of an SC lattice polygon is equal to $\pi/2$ times
the number of right angles between two edges.  The average
total curvature of minimal length polygons of knot type $K$
is denoted in units of $2\pi$ by $\C{K}_K$ (that is, the
average total curvature is $2\pi\C{K}_K$).  Obviously 
$\C{K}_{0_1} = 1$ in the SC lattice, since every minimal length
unknotted polygon of length $4$ is a unit square of total curvature
$2\pi$.  For other knot types the total curvature of a polygon is
an integer multiple of $\pi/2$, and the mean curvature is thus a
rational number times $2\pi$.  Hence, for a knot type $K$,
the average curvature of minimal length length polygons of knot type
$K$ is given by
\begin{equation}
\LL C_K \RR = 2\pi \C{K}_K
\label{eqn21A} %%ZXZ[eqn21A]
\end{equation}
where $\C{K}_K$ is a rational number. 

Similar definitions hold for polygons in the BCC and FCC.  In each
case the lattice curvature of a polygon is the sum of the complements
of angles inscribed between successive edges.

In the FCC the curvature of a polygon is a summation over angles
of sizes $0$, $\pi/3$ and $2\pi/3$.  Hence $2\pi \C{K}_K$ is
a rational number similarly to the case in the SC.  This gives
a similar definition to equation \Ref{eqn21A} of $\C{K}_K$ for
minimal length lattice polygons of knot type $K$.  Obviously, 
$\C{K}_{0_1} = 1$ in the FCC, since each minimal lattice polygon
of knot type $0_1$ is an elementary equilateral triangle.

The situation is somewhat more complex in the BCC lattice.  The 
curvature of a polygon is the sum over angles of sizes 
$\arccos (1/\sqrt{3})$, $\pi - \arccos (1/\sqrt{3})$ and $0$.  
This shows that the average curvature of minimal length polygons 
of knot type $K$ is of the generic form
\begin{equation}
\LL C_K \RR = \C{B}_K \arccos (1/\sqrt{3}) + 2\pi \C{K}_K 
\label{eqn21B} %%ZXZ[eqn21B]
\end{equation}
where $\C{B}_K$ and $\C{K}_K$ are rational numbers.  By examining
the $12$ minimal length unknotted polygons of length $4$ in the
BCC, one can show that $\C{B}_{0_1} = -2$ and $\C{K}_{0_1} = 3/2$.

The minimal lattice curvature  $\C{C}_K$ (as opposed to the 
average curvature) of SC lattice knots were examined in reference 
\cite{JvRP99}.\footnote{Observe that the minimal lattice curvature 
of a lattice knot does not necessarily occur at minimal length.}
For example, it is known that $\C{C}_{0_1} = 2\pi$ while $\C{C}_{3_1} 
= 6\pi$ in the SC lattice \cite{JvRP99}.  Bounds on the minimal 
lattice curvature in the SC lattice can also be found in terms of 
the minimal crossing number $C_K$ or the bridge number $b_K$ of a knot.  
In particular, $\C{C}_{K} \geq \max\LC \LB 3+\sqrt{9+8C_K}\RB \pi/ 4 
\, , \, 3\pi b_K \RC$.  These bounds are in particular good enough 
to prove that $\C{C}_{9_{47}} = 9\pi$. A minimal lattice curvature 
index $\nu_K$ is also proven to exist in reference \cite{JvRP99}, 
in particular
\begin{equation}
\lim_{n\to\infty} \frac{\C{C}_{K^n}}{n} = \nu_K
\label{eqn25} %%ZXZ[eqn25]
\end{equation}
exists and $\C{C}_{K^n} \geq n \nu_K$.  It is known that 
$\nu_{0_1} = 0$ but that $2\pi \leq \nu_{3_1} \leq 3\pi$ in the 
SC lattice, and one expect that $2\pi\C{K}_{K^n} \geq \C{C}_{K^n}
\geq 2\pi n$ in the SC lattice. This shows that $\C{K}_{K^n}$
increases at least as fast as $n$ in the SC lattice.  
For more details, see reference \cite{JvRP99}.
 
%%%%%%%%%%%%%%%%%%%%%%%%%%%%%%%%%%%%%%%%%%%%%%%%%%%%%%%%%%%%%%%%%%
%%%%%%%%%%%%%%%%%%%%%%%%%%%%%%%%%%%%%%%%%%%%%%%%%%%%%%%%%%%%%%%%%%
\section{GAS Sampling of knotted polygons}

Knotted polygons can be sampled by implementing the GAS
algorithm \cite{JvRR09}.  The algorithm is implemented using
a set of local elementary transitions (called ``atmospheric
moves" \cite{JvRR08}) to sample along sequences of polygon
conformations. The algorithm is a generalisation of 
the Rosenbluth algorithm \cite{RR55}, and is an approximate 
enumeration algorithm \cite{JvR09,JvR10}.

The GAS algorithm can be implemented in the SC lattice
on polygons of given knot type $K$ using the BFACF elementary
moves \cite{AC83,ACF83,BF81} to implement the atmospheric moves
\cite{JvRR10,JvRR11}.  These elementary moves are illustrated
in figure \ref{fig2}. This implementation is irreducible on classes
of polygon of fixed knot type \cite{JvRW91}.

%%%%%%%%%%%%%%%%%
\begin{figure}[t]
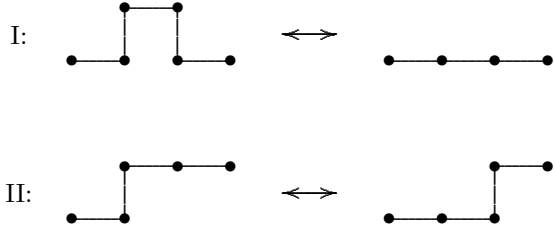

\input fig3.inp
\caption{BFACF elementary moves on polygons in the cubic lattice.
These (reversible) moves are of two types:  Type I decreases
or increases the length of the polygon by two edges, while
Type II is a neutral move which maintains the length of the
polygon.  A move which increases the length of the polygon
is a positive move, while negative moves decrease the length
of the polygon.}
\label{fig3} %%ZXZ[fig3]
\end{figure}
%%%%%%%%%%%%%%%%

The BFACF moves in figure \ref{fig3} are either positive
(increase the length of a polygon), neutral (leave the length
unchanged) or negative (decrease the length of a polygon).
These moves define the atmosphere of a polygon.  The collection
of possible positive moves constitutes the \textit{positive atmosphere}
of the polygon.  Similarly, the collection of neutral moves
composes the \textit{neutral atmosphere} while the set of negative
moves is the \textit{negative atmosphere} of the polygon. The
the size of an atmosphere of a polygon $\omega$ is the number of
possible successful elementary moves that can be performed to change
it into a different conformation.  We denote the size of the positive
atmosphere of a polygon $\omega$ by $a_+(\omega)$, of the 
neutral atmosphere by $a_0(\omega)$, and of the negative atmosphere 
by $a_-(\omega)$.

The GAS algorithm is implemented on cubic lattice polygons
as follows (for more detail, see references \cite{JvRR10,JvRR11}).
Let $\omega_0$ be a lattice polygon of knot type $K$, then 
sample along a sequence of polygons $\LL\omega_0, \omega_1,
\omega_2, \ldots \RR$ by updating $\omega_i$ to $\omega_{i+1}$
using an atmospheric move.

Each atmospheric move is chosen uniformly from the collection
of possible moves in the atmospheres.  That is, if $\omega_j$ has 
length $\ell_j$ then the probabilities for positive, neutral 
and negative moves are given by
\begin{equation}
\Pr(\hbox{$+$}) \propto \beta_{\ell_j} a_+(\omega_j),\quad
\Pr(\hbox{$0$}) \propto a_0(\omega_j),\quad\hbox{and}\,
\Pr(\hbox{$-$}) \propto a_-(\omega_j) 
\end{equation}
where the parameters $\beta_{\ell}$ were introduced in order to
control the transition probabilities in the algorithm.  It will 
be set in the simulation for ``flat sampling".  That, it will
be chosen approximately equal to the ratio of average sizes of 
the positive and negative atmospheres of polygons of length $\ell$: 
$\beta_\ell \approx \frac{\mean{a_+}_{\ell} }{ \mean{a_-}_{\ell} }$.  
This choice makes the average probability of a positive atmospheric 
move roughly equal to the probability of a negative move at each value 
of $\ell$.

%%%%%%%%%%%%%%%%%
\begin{figure}[b]
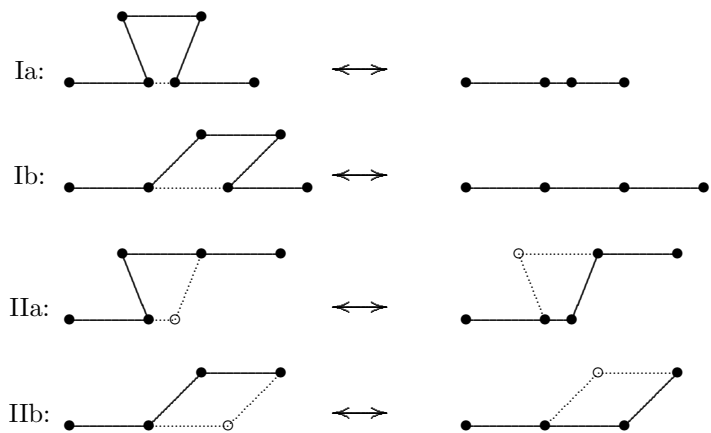

\input fig4.inp
\caption{Elementary moves on polygons in the BCC lattice.
These (reversible) moves are of two types:  Type I decreases
or increases the length of the polygon by two edges, while
Type II is a neutral move which maintains the length of the
polygon.}
\label{fig4} %%ZXZ[fig4]
\end{figure}
%%%%%%%%%%%%%%%%

%%%%%%%%%%%%%%%%%
\begin{figure}[t]
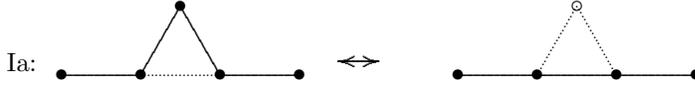

\input fig5.inp
\caption{The Elementary move on polygons in the FCC lattice.
This is the only class of elementary moves in this lattice,
there are no neutral moves.}
\label{fig5} %%ZXZ[fig5]
\end{figure}
%%%%%%%%%%%%%%%%

This sampling produces a sequence $\langle\omega_j \rangle$ 
of states and we assign a weight 
\begin{equation}
W(\omega_n) = \LH
\frac{a_-(\omega_0) + a_0(\omega_0) + \beta_{\ell_0} a_+(\omega_0) }
{a_-(\omega_{n}) + a_0(\omega_{n}) + \beta_{\ell_{n}} a_+(\omega_{n}) }\RH 
\times \prod_{j=0}^n \beta_{\ell_j}^{(\ell_j - \ell_{j+1})} .
\end{equation}
to the state $\omega_n$.  The GAS algorithm is an approximate enumeration
algorithm in the sense that the ratio of average weights of polygons
of lengths $n$ and $m$ tends to the ratio of numbers of such polygons.
That is, 
\begin{equation}
  \frac{\mean{W}_n}{\mean{W}_m} = \frac{p_n(K)}{p_m(K)}.
  \label{eqn w ratio} %%ZXZ[eqn w ratio]
\end{equation}
The algorithm was coded using hash-coding such that updates of 
polygons and polygon atmospheres were done in $O(1)$ CPU time.  
This implementation was very efficient, enabling us to perform 
billions of iterations on knotted polygons in reasonable real 
time on desk top linux workstations.  Minimal length polygons 
of each knot type were sieved from the data stream and hashed 
in a table to avoid duplicate discoveries.  The lists of minimal 
length polygons were analysed separately  by counting symmetry 
classes, and computing writhes and curvatures.

Implementation of GAS sampling in the FCC and BCC lattices proceeds
similar to the implementation in the SC lattice.   It is only
required to define suitable atmospheric moves analogous to the
SC lattice moves in figure \ref{fig3}, and to show that these moves
are irreducible on classes of FCC or BCC lattice polygons of fixed
knot types. 

The BCC lattice has girth four, and local positive, neutral and negative
atmospheric moves similar to the SC lattice moves in figure \ref{fig3}
can be defined in a very natural way.  These are illustrated in
figure \ref{fig4}.  Observe that the conformations in this figure
are not necessarily planar, in particular because minimal length
lattice polygons in the BCC lattice are not necessarily planar.
This collection of elementary moves is irreducible on classes of
unrooted lattice polygons of fixed knot type $K$ in the BCC lattice
\cite{JvRR11,JvRR11a}.

In the FCC lattice the generalisation of the BFACF elementary moves 
is a single class of positive atmospheric moves and their inverse,
illustrated in figure \ref{fig5}.  This elementary move (and its
inverse) is irreducible on classes of unrooted lattice polygons of 
fixed knot type $K$ in the FCC lattice \cite{JvRR11}. The 
implementation of this elementary move using the GAS algorithm is
described in references \cite{JvRR10,JvRR11,JvRR11a}.

%%%%%%%%%%%%%%%%%%%%%%%%%%%%%%%%%%%%%%%%%%%%%%%%%%%%%%%%%%%%%%%%%%
%%%%%%%%%%%%%%%%%%%%%%%%%%%%%%%%%%%%%%%%%%%%%%%%%%%%%%%%%%%%%%%%%%
%%%%%%%%%%%%%%%%%%%%%%%%%%%%%%%%%%%%%%%%%%%%%%%%%%%%%%%%%%%%%%%%%%
\section{Numerical Results}

GAS algorithms for knotted polygons in the SC, BCC and FCC lattices
were coded and run for polygons of lengths $n\leq M$ where
$500 \leq M \leq 700$, depending on the knot type (the larger
values of $M$ were used for more complicated compound knots).  In
each simulation, up to 500 GAS sequences each of length $10^7$ states
were realised with the purpose of counting and collecting minimal 
length polygons.  In most cases the algorithm efficiently 
found minimal conformations in short real time, but a few knots
proved problematic, and in particular compound knots.  For
example, knots types $(3_1^+)^5$ and $(4_1)^3$ required weeks
of CPU time, while $(3_1^+)^2(3_1^-)^2$ proved to be beyond the
memory capacity of our computers.

Generally, our simulations produced lists of symmetry classes
of minimal length knotted polygons in the three lattices.
Our data (lists of minimal length knotted polygons) are 
available at the website in reference \cite{PolyKnots}.

%%%%%%%%%%%%%%%%%%%%%%%%%%%%%%%%%%%%%%%%%%%%%%%%%%%
%%%%%%%%%%%%%%%%%%%%%%%%%%%%%%%%%%%%%%%%%%%%%%%%%%%
\subsection{Minimal Knots in the Simple Cubic Lattice}

%%%%%%%%%%%%%%%%%%%%%%%%%%%%%%%%%%%%%%%%%%%%%%%%%%%%%%%
\subsubsection{Minimal Length SC Lattice Knots:}
The minimal lengths $n_K$ of prime knot types $K$ are displayed
in table \ref{tablenKSC}.  We limited our simulations to prime
knots up to eight crossings.  In addition, a few knots with more
than eight crossings were included in the table, including
the first two knots in the knot tables to 12 crossings, as
well as $9_{42}$ and $9_{47}$. The minimal lengths of some
compound knots (up to eight crossings), as well as compound
trefoils up to $(3_1^+)^6$ and figure eights up to $(4_1)^6$,
were also examined, and data are displayed in table \ref{tablenKSCC}.

%%%%%%%%%%%%%%%%%XXXXXXXXXXXXXXXXXXXXXX
\begin{table}[t!]
\begin{center}
 \begin{tabular}{||c|| l ||}
 \hline
  $n_K$ & Prime Knot types  \\
 \hline 
$4$ & $0_1$  \\ 
$24$ & $3_1$  \\
$30$ & $4_1$  \\
$34$ & $5_1$  \\
$36$ & $5_2$  \\
$40$ & $6_1$, $6_2$, $6_3$ \\
$42$ & $8_{19}$  \\
$44$ & $7_1$, $7_3$, $7_4$, $7_7$, $8_{20}$  \\
$46$ & $7_2$, $7_5$, $7_6$, $8_{21}$ \\
$48$ & $8_3$, $8_7$, $9_{42}$  \\
$50$ & $8_1$, $8_2$, $8_4$, $8_5$, $8_6$, $8_8$, $8_9$, $8_{10}$,
       $8_{11}$, $8_{13}$, $8_{14}$, $8_{16}$, $9_{47}$ \\
$52$ & $8_{12}$, $8_{15}$, $8_{17}$, $8_{18}$  \\
$54$ & $9_1$  \\
$56$ & $9_2$  \\
$60$ & $10_1$, $10_2$  \\
$64$ & $11_1$  \\
$66$ & $11_2$  \\
$70$ & $12_1$, $12_2$  \\
 \hline
 \end{tabular}
\end{center}
 \caption{Minimal Length of Prime Knots in the SC lattice.}
  \label{tablenKSC}
\end{table}
%%%%%%%%%%%%%%%%%XXXXXXXXXXXXXXXXXXXXXX

The results in tables \ref{tablenKSC} and \ref{tablenKSCC} confirms
data previously obtained for minimal knots in the simple cubic lattice,
see for example \cite{JvR99} and in particular reference \cite{DIASV09}
for extensive results on minimal length knotted SC polygons. 

The number of different knot types with minimal length $n_k \leq n$
can be estimated and grows exponentially with $n$.
In fact, if $Q_n$ is the number of different knot types with 
$n_K \leq n$, then $Q_m \geq Q_n$ if $m\geq n$.  Obviously, $Q_n \leq
p_n$, so that 
\begin{equation}
\C{Q} = \limsup_{n\to\infty} Q_n^{1/n} \leq \lim_{n\to\infty} p_n^{1/n}
= \mu
\end{equation}
by equation \Ref{eqn1AA}.

On the other hand, suppose that $N>0$ prime knot types (different
from the unknot) can be tied in polygons of length $m$ (that is 
$Q_m \geq N$).  Then by concatenating $k$ polygons of different prime 
knot types as in figure \ref{fig1}, it follows that $Q_{Nm} \geq 
\sum_{k=0}^N \LB {{N}\atop{k}} \RB = 2^N$.  In other words
\begin{equation}
Q_{n} \geq 2^{n/m} .
\end{equation}
if $n=Nm$ where $N$ is the number of non-trivial prime knot types
that can be tied in a polygon of length $n$.  For example, if $m=28$,
then $N=1$, and thus $Q_{28} \geq 2$.  Taking $n\to\infty$ implies
that $N\to\infty$ as well so that 
$\liminf_{n\to\infty} Q_n^{1/n} \geq 1$. 

In other words, $1\leq \C{Q} \leq \mu$.

Thence, one may estimate $Q_n^{1/n}$, and increasing $n$ 
in $Q_n^{1/n}$ should give increasingly better estimates of $\C{Q}$. 
In addition, if $Q_n^{1/n}$ approaches a limit bigger than one, then 
$\C{Q}>1$ and the number of different knot types that can be tied 
in a polygon of length $n$ increases exponentially with $n$.

By examining the data in tables \ref{tablenKSC} and \ref{tablenKSCC},
one observes that $Q_{30} = 4$ so that $\C{Q} \approx 4^{1/30}
\approx  1.0472\ldots$.  Increasing $n$ to $40$ gives
$Q_{40}=16$, so that $\C{Q} \approx 1.07177\ldots$.  If
$n=50$, then $Q_{50} \geq 74$, hence $\C{Q} \approx 1.08989\ldots$. 
These approximate estimates of $\C{Q}$ increases systematically,
suggesting the estimates are lower bounds, and that $\C{Q}>1$.

The number of distinct knot types with $n_K = n$ is 
$\ol{Q}_n = Q_{n} - Q_{n-1}$, and since $Q_n \geq Q_{n-1}$ and
$Q_n = \C{Q}^{n+o(n)}$, it follows that $\ol{Q}_n = 0$ if 
$n$ is odd, and $\ol{Q}_n = \C{Q}^{n+o(n)}$ for even values of $n$.

There appears to be several cases of regularity amongst the 
minimal lengths of knot types in tables \ref{tablenKSC}.
The sequence of $(N,2)$-torus knots with
$N\geq 3$ (these are the knots $\{3_1,5_1,7_1,9_1,11_1\}$)
increases in steps of $10$ starting in $24$.  Similarly, 
the sequence of twist knots $\{4_1,6_1,8_1,10_1,12_1\}$ 
increments in $10$ starting in $30$, as do the sequence
of twist knots $\{5_2,7_2,9_2,11_2\}$, but starting in $36$.
The sequence $\{4_1,6_2,8_2,10_2,12_2\}$ also increments
in $10$, starting at $30$ as well. A discussion of
these patterns can be found in reference \cite{JvR99} (see
figure 3 therein).  There are no proofs that these patterns 
will persist indefinitely.

%%%%%%%%%%%%%%%%%XXXXXXXXXXXXXXXXXXXXXX
\begin{table}[t!]
\begin{center}
 \begin{tabular}{||c|| l ||}
 \hline
  $n_K$ & Compound Knot Types \\
 \hline 
$40$ & $(3_1^+)^2$, $(3_1^+)(3_1^-)$ \\
$46$ & $(3_1^+)(4_1)$ \\
$50$ & $(5_1^+)(3_1^+)$, $(5_1^+)(3_1^-)$, $(5_2^+)(3_1^-)$ \\
$52$ & $(4_1)^2$, $(5_2^+)(3_1^+)$ \\
$56$ & $(3_1^+)^3$, $(3_1^+)^2(3_1^-)$ \\
$72$ & $(3_1^+)^4$ \\
$74$ & $(4_1^+)^3$ \\
$88$ & $(3_1^+)^5$ \\
$96$ & $(4_1)^4$ \\
$104$ & $(3_1^+)^6$ \\
$118$ & $(4_1)^5$ \\
$140$ & $(4_1)^6$ \\
 \hline
 \end{tabular}
\end{center}
 \caption{Minimal Length of Compound Knots in the SC lattice.}
  \label{tablenKSCC}
\end{table}
%%%%%%%%%%%%%%%%%XXXXXXXXXXXXXXXXXXXXXX

In table \ref{tablenKSCC} the estimated minimal lengths $n_K$ 
of a few compounded knots are given.  These data similarly
exhibit some level of regularity.  For example, the family
of compounded positive trefoils $\LC (3_1^+)^n \RC$ increases in
steps of $16$ starting in $24$. From these data, one may bound 
the minimal lattice edge index of positive trefoils (defined
in equation \Ref{eqn10}).  In particular, $\alpha_K \leq
n_{K^p}/p$, and if $K=3_1^+$ and $p=6$, then it follows that
$\alpha_{3_1^+} \leq 17\frac{1}{3}$.  This does not improve on
the upper bound given in references \cite{JvRP95} and \cite{JvR96},
but if the increment of $16$ persists, then if $p=10$ one would 
obtain $\alpha_{3_1^+} \leq 16\frac{4}{5} < 17$.  Preliminary 
calculations indicated that finding the minimal edge number
for $(3_1^+)^{10}$ would be a difficult simulation, and this 
was not pursued.  At this point, the argument illustrated in
figure 4 in reference \cite{JvR99} proves that $\alpha_{3_1^+} \leq 17$,
and the data above suggest that $\alpha_{3_1} = 24 + 16n$ for 
$n \leq 6$.  If this pattern persists, then $\alpha_K$ would
be equal to $16$, but there is no firm theoretical argument
which validates this expectation.

Similar observations apply to the family of compounded
figure eight knots $\LC (4_1)^n \RC$. The minimal edge numbers
for $n\leq 6$ are displayed in table \ref{tablenKSCC} and
increments by $22$ such that $\alpha_{4_1} = 30 + 22n$ for $n \leq 6$.
This suggest that $\alpha_{4_1} = 22$, but the best upper bound
from the data in table \ref{tablenKSCC} is $23\frac{2}{3}$.

%%%%%%%%%%%%%%%%%XXXXXXXXXXXXXXXXXXXXXX
\begin{table}[t!]
\begin{center}
 \begin{tabular}{||c||r|r|r|r||l|l|l||}
 \hline
  Knot & \multicolumn{7}{|c||}{Simple Cubic Lattice} \\
  \hline  
   & $n_K$ & $\C{P}_K$ & \multicolumn{2}{|c||}{$\C{S}_K$} 
    & $\C{W}_K$ & $\C{|W|}_K$ & $\C{K}_K$ \\
   \hline
$0_1$ & 4  & 3     & 1   & $3^1$ & \SMALL 0  & \SMALL 0 
    & \SMALL 1 \\
$3_1$ & 24 & 3328  & 142  & $24^{137}8^5$    
    &  \SMALL $3\frac{735}{1664}$ &  \SMALL $3\frac{735}{1664}$ 
    &  \SMALL $3 \frac{801}{1664}$ \\
$4_1$ & 30 & 3648  & 152 & $24^{152}$ 
    &  \SMALL $0$ &  \SMALL $\frac{33}{152}$ 
    &  \SMALL $4 \frac{1}{152}$ \\
$5_1$ & 34 & 6672  & 278 & $24^{278}$
    &  \SMALL $6\frac{127}{556}$ &  \SMALL $6\frac{127}{556}$ 
    & \SMALL $4\frac{459}{556}$ \\
$5_2$ & 36 & 114912 & 4788 & $24^{4788}$
    &  \SMALL $4\frac{5057}{9576}$ &  \SMALL $4\frac{5057}{9576}$ 
    & \SMALL $5\frac{61}{9576}$ \\
$6_1$ & 40 & 6144  & 258 & $24^{254}12^4$
    &  \SMALL $1\frac{49}{512}$ &  \SMALL $1\frac{49}{512}$ 
    & \SMALL $5\frac{223}{512}$ \\
$6_2$ & 40 & 32832 & 1368 & $24^{1368}$
    &  \SMALL $2\frac{1079}{1368}$ &  \SMALL $2\frac{1079}{1368}$ 
    & \SMALL $5\frac{65}{456}$ \\
$6_3$ & 40 & 3552  & 148 & $24^{148}$
    &  \SMALL $0$ &  \SMALL $\frac{29}{148}$ 
    & \SMALL $4\frac{36}{37}$ \\
$7_1$ & 44 & 33960  & 1415 & $24^{1415}$
    &  \SMALL $9\frac{61}{283}$ &  \SMALL $9\frac{61}{283}$ 
    & \SMALL $6\frac{322}{1415}$ \\
$7_2$ & 46 & 336360  & 14016 & $24^{14014}12^2$
    &  \SMALL $5\frac{10539}{14015}$ 
    &  \SMALL $5\frac{10539}{14015}$ 
    & \SMALL $6\frac{811}{28030}$ \\
$7_3$ & 44 & 480  & 20 & $24^{20}$
    &  \SMALL $7\frac{19}{40}$ &  \SMALL $7\frac{19}{40}$ 
    & \SMALL $5\frac{31}{40}$ \\
$7_4$ & 44 & 168  & 7 & $24^{7}$
    &  \SMALL $5\frac{4}{7}$ &  \SMALL $5\frac{4}{7}$ 
    & \SMALL $5\frac{2}{7}$ \\
$7_5$ & 46 & 9456  & 394 & $24^{394}$
    &  \SMALL $7\frac{107}{394}$ &  \SMALL $7\frac{107}{394}$ 
    & \SMALL $6\frac{83}{394}$ \\
$7_6$ & 46 & 34032  & 1418 & $24^{1418}$
    &  \SMALL $3\frac{625}{1418}$ 
    &  \SMALL $3\frac{625}{1418}$ & \SMALL $5\frac{614}{709}$ \\
$7_7$ & 44 & 504  & 21 & $24^{21}$
    &  \SMALL $\frac{43}{84}$ &  \SMALL $\frac{43}{84}$ 
    & \SMALL $5\frac{1}{12}$ \\
$8_1$ & 50 & 23736 & 990 & $24^{988}12^2$
    &  \SMALL $2\frac{813}{1978}$ &  \SMALL $2\frac{813}{1978}$ 
    & \SMALL $6\frac{514}{989}$ \\
$8_2$ & 50 & 91680 & 3820 & $24^{3820}$
    &  \SMALL $5\frac{2081}{3820}$ &  \SMALL $5\frac{2081}{3820}$ 
    & \SMALL $6\frac{1259}{3820}$ \\
$8_3$ & 48 & 12 & 1 & $12^{1}$
    &  \SMALL $0$ &  \SMALL $0$ 
    & \SMALL $5\frac{91}{97}$ \\
$8_4$ & 50 & 47856 & 1994 & $24^{1994}$
    &  \SMALL $1\frac{2613}{3988}$ &  \SMALL $1\frac{2613}{3988}$ 
    & \SMALL $6\frac{1675}{3988}$ \\
$8_5$ & 50 & 1152 & 48 & $24^{48}$
    &  \SMALL $5\frac{5}{8}$ &  \SMALL $5\frac{5}{8}$  
    & \SMALL $6\frac{5}{24}$ \\
$8_6$ & 50 & 11040 & 460 & $24^{460}$
    &  \SMALL $4\frac{7}{920}$ &  \SMALL $4\frac{7}{920}$  
    & \SMALL $6\frac{273}{920}$ \\
$8_7$ & 48 & 48 & 2 & $24^{2}$
    &  \SMALL $2\frac{3}{4}$ &  \SMALL $2\frac{3}{4}$  
    & \SMALL $5\frac{3}{4}$ \\
$8_8$ & 50 & 3120 & 130 & $24^{130}$
    &  \SMALL $1\frac{21}{260}$ &  \SMALL $1\frac{21}{260}$  
    & \SMALL $6\frac{89}{260}$ \\
$8_9$ & 50 & 35280 & 1470 & $24^{1470}$
    &  \SMALL $0$ &  \SMALL $\frac{527}{2940}$  
    & \SMALL $6\frac{307}{980}$ \\
$8_{10}$ & 50 & 1680 & 70 & $24^{70}$
    &  \SMALL $3\frac{1}{140}$ &  \SMALL $3\frac{1}{140}$  
    & \SMALL $5\frac{121}{140}$ \\
$8_{11}$ & 50 & 192 & 8 & $24^{8}$
    &  \SMALL $4\frac{1}{8}$ &  \SMALL $4\frac{1}{8}$  
    & \SMALL $6\frac{3}{8}$ \\
$8_{12}$ & 52 & 2592 & 108 & $24^{108}$
    &  \SMALL $\frac{71}{216}$ &  \SMALL $\frac{71}{216}$  
    & \SMALL $6\frac{187}{216}$ \\
$8_{13}$ & 50 & 26112 & 1088 & $24^{1088}$
    &  \SMALL $1\frac{399}{2176}$ &  \SMALL $1\frac{399}{2176}$  
    & \SMALL $5\frac{2099}{2176}$ \\
$8_{14}$ & 50 & 720 & 30 & $24^{30}$
    &  \SMALL $3\frac{59}{60}$ &  \SMALL $3\frac{59}{60}$  
    & \SMALL $6\frac{29}{60}$ \\
$8_{15}$ & 52 & 80208 & 3342 & $24^{3342}$
    &  \SMALL $7\frac{2957}{3342}$ 
    &  \SMALL $7\frac{2957}{3342}$ & \SMALL $6\frac{549}{1114}$ \\
$8_{16}$ & 50 & 96 & 4 & $24^{4}$
    &  \SMALL $2\frac{5}{8}$ &  \SMALL $2\frac{5}{8}$ 
    & \SMALL $5\frac{7}{8}$ \\
$8_{17}$ & 52 & 53184 & 2216 & $24^{2216}$
    &  \SMALL $0$ &  \SMALL $\frac{1099}{4432}$ 
    & \SMALL $6\frac{321}{4432}$ \\
$8_{18}$ & 52 & 3552 & 148 & $24^{148}$
    &  \SMALL $0$ &  \SMALL $\frac{10}{37}$ 
    & \SMALL $5\frac{71}{148}$ \\
$8_{19}$ & 42 & 13992 & 592 & $24^{574}12^{18}$
    &  \SMALL $8\frac{885}{1166}$ 
    &  \SMALL $8\frac{885}{1166}$ & \SMALL $5\frac{267}{583}$ \\
$8_{20}$ & 44 & 240 & 10 & $24^{10}$
    &  \SMALL $2\frac{1}{10}$ &  \SMALL $2\frac{1}{10}$ 
    & \SMALL $4\frac{9}{10}$ \\
$8_{21}$ & 46 & 56040 & 2335 & $24^{2335}$
    &  \SMALL $4\frac{2499}{4670}$ &  \SMALL $4\frac{2499}{4670}$ 
    & \SMALL $5\frac{373}{467}$ \\
$9_1$ & 54 & 345960 & 14417 & $24^{14414}8^3$
    &  \SMALL $12\frac{283}{4805}$ &  \SMALL $12\frac{283}{4805}$
    & \SMALL $7\frac{5763}{9610}$ \\
$9_2$ & 56 & 3281304 & 136721 & $24^{136721}$
    &  \SMALL $6\frac{105057}{136721}$ 
    &  \SMALL $6\frac{105057}{136721}$ 
    & \SMALL $7\frac{96676}{136721}$ \\
$9_{42}$ & 48 & 27744 & 1156 & $24^{1156}$
    &  \SMALL $1\frac{953}{2312}$ 
    &  \SMALL $1\frac{953}{2312}$ 
    & \SMALL $5\frac{1939}{2312}$ \\
$9_{47}$ & 50 & 13680 & 570 & $24^{570}$
    &  \SMALL $2\frac{4}{5}$ 
    &  \SMALL $2\frac{4}{5}$ 
    & \SMALL $5\frac{47}{95}$ \\
$10_1$ & 60 & 462576 & 19298 & $24^{19250}12^{48}$
    &  \SMALL $3\frac{12507}{38548}$ 
    &  \SMALL $3\frac{12507}{38548}$ 
    & \SMALL $8\frac{5341}{38548}$ \\
$10_2$ & 60 & 871296 & 36304 & $24^{36304}$
    &  \SMALL $8\frac{42761}{72608}$ 
    &  \SMALL $8\frac{42761}{72608}$ 
    & \SMALL $7\frac{56813}{72608}$ \\
\hline
 \end{tabular}
\end{center}
 \caption{Data on prime knot types in the SC Lattice.}
  \label{tableSC}
\end{table}
%%%%%%%%%%%%%%%%%XXXXXXXXXXXXXXXXXXXXXX

%%%%%%%%%%%%%%%%%%%%%%%%%%%%%%%%%%%%%%%%%%%%%%%%%%%%%%%%%%%%%%%%%
\subsubsection{Entropy of minimal lattice knots in the SC lattice:}

Minimal length lattice knots were sieved from the data stream,
then classified and stored during the simulations, which was allowed 
to continue until all, or almost all, minimal length lattice were
discovered. In several cases a simulation was repeated in order
to check the results. We are very sure of our data if 
$\C{P}_K \lesssim 1000$, reasonable certain if 
$1000 \lesssim \C{P}_K \lesssim 10000$, less certain if 
$10000\lesssim \C{P}_K \lesssim 100000$, and we consider 
the stated value of $\C{P}_K$ to be only a lower
bound if $\C{P}_K \gtrsim 100000$ in table \ref{tableSC}.

Data on entropy, lattice writhe and lattice curvature, were collected
on prime knot types up to eight crossings, and also the knot types
$9_1$, $9_1$, $9_{42}$, $9_{47}$, $10_1$ and $10_2$.  The
SC lattice data are displayed in table \ref{tableSC}.  As before,
the minimal length of a knot type $K$ is denoted by $n_K$,
and $\C{P}_K = c_{n_K}(K)$ is the total number of minimal length
SC lattice knots of length $n_K$.  For example, there are $3328$
minimal length trefoils (of both chiralities) of length $n_{3_1}=24$.
Since $3_1$ is chiral, $\C{P}_{3_1^+} = 3328/2 = 1664$.  

The unknot has minimal length $4$, which is a unit square polygon 
in a symmetry class of $3$ members which are equivalent under 
lattice symmetries.  The $3328$ minimal lattice trefoils 
are similarly partitioned into $142$ symmetry classes, of which 
$137$ classes has $24$ members and $5$ classes has $8$ members 
each.  These partitionings into symmetry classes are denoted by 
$3^1$ for the unknot, and $24^{137}8^5$ for lattice polygons of 
knot type $3_1$ (of both chiralities) or $12^{137}4^5$ for 
lattice polygons of (say) right-handed knot type $3_1^+$.

Entropy per unit length of minimal polygons of knot type $K$
is defined by
\begin{equation}
\C{E}_K = \frac{\log \C{P}_K}{n_K}  .
\label{eqn31} %%ZXZ[eqn31]
\end{equation}
This is a measure of the tightness of the minimal knot.  If
$\C{E}_K$ is small, then there are few conformations that the
minimal knot can explore, and such a knot is tightly embedded
in the lattice (and its edges are relatively immobile).  If 
$\C{E}_K$, on the other hand, is large, then there are a relatively
large conformational space which the edges may explore, and such
a knot type is said to be loosely embedded.

The unknot has $\C{E}_{0_1} = (\log 3)/4 = 0.27467\ldots$, which
will be small compared to other knot types, and is thus tightly
embedded.

The entropy per unit length of the (right-handed) trefoils
is $\C{E}_{3_1^+} = (\log  1664)/24 = 0.3090\ldots$, and it appears
that the edges in these tight embeddings are similarly constrained
to those in the unknot.  Edges in the (achiral) knot $4_1$ has
$\C{E}_{4_1} = (\log 3648)/30 = 0.2733\ldots < \C{E}_{3_1^+}$, and are 
more constrained than those in the trefoil.  Similarly, for 
five crossing knots one finds that $\C{E}_{5_1^+} = 0.2386\ldots$
while $\C{E}_{5_2^+} = 0.3223\ldots$.

The entropy per unit length seems to converge in families of
knot types.  For example in the $(N,2)$-torus knot 
family $\{3_1^+,5_1^+,7_1^+,9_1^+\}$ one gets 
$\{0.3090,0.2386,0.2214,0.2234\}$ to four digits accuracy.
Similarly, the family of twist knots $\{3_1^+,5_2^+,7_2^+,9_2^+\}$
gives $\{0.3090,0.3044,0.2616,0.2555\}$, again to four digits.
Similar patterns are observed for the families $\{6_1^+,8_1^+,
10_1^+\}$ ($\{0.2008,0.1876,0.2059\}$), and $\{6_2^+,8_2^+,10_2^+\}$
($\{0.2427,0.2147,0.2164\}$).  Further extensions of the 
estimates of $\C{P}_K$ for more complicated knots would be necessary
to test these patterns, but the scope of such simulations are beyond
our available computing resources.

Finally, there are some knots with very low entropy per unit length.
These include $7_3^+$ ($0.1246$), $7_4^+$ ($0.1007$), $7_7^+$
($0.1257$), $8_3$ ($0.08065$), $8_7^+$ ($0.06621$), 
$8_{11}^+$ ($0.09129$), $8_{16}^+$ ($0.07742$) and 
$8_{20}^+$ ($0.1088$).  These knots are tightly embedded in the SC
lattice in their minimal conformations, with very little entropy 
per edge available.

The distribution of minimal knotted polygons in symmetry classes
in table \ref{tableSC} shows that most minimal knotted polygons
are not symmetric with respect to elements of the octahedral 
group, and thus fall into classes of $24$ distinct polygons.
Classes with fewer elements, (for example $12$ or $8$), has
symmetric embeddings of the embedded polygons.  Such
symmetric embeddings are the exception rather than the rule in
table \ref{tableSC}:  For example, amongst the listed prime
knot types in that table, only eight types admit to a symmetric
embedding.

%%%%%%%%%%%%%%%%%%%%%%%%%%%%%%%%%%%%%%%%%%%%%%%%%%%%%
\subsubsection{The Lattice Writhe and Curvature:}

The average writhe $\C{W}_K$, the average absolute writhe $|\C{W}|_K$
and the average curvature $\C{K}_K$ (in units of $2\pi$)
of minimal length polygons are displayed in table \ref{tableSC}.  
The results are given as rational numbers, since these numbers 
can be determined exactly from the data. Observe that the writhe
of simple cubic lattice polygons are known to be rational numbers
\cite{LS91,JvROSTW96,JvRSW99}, hence the average over finite sets 
of polygons will also be rational.  In addition, the average
writhe $\C{W}_K$ is non-negative in table \ref{tableSC} since 
the right handed knot was in each case used in the simulation.

In most cases in table \ref{tableSC} it was observed that
$|\C{W}_K| = |\C{W}|_K$, with the exception of some achiral knots,
which have $|\C{W}|_K>0$ while $\C{W}_K=0$.  The average absolute
writhe was zero in only two cases, namely the unknot and the
knot $8_3$.  Generally, the average and absolute average writhe
of achiral knots are not equal, but the unknot and $8_3$ are
exceptions to this rule.  

%%%%%%%%%%%%%%%%%%%%%%%%%%%%%%%%%%%%%%%%%%%%%
\begin{table}[t!]
\begin{center}
 \begin{tabular}{||c||r|r|r|r||l|l|l||}
 \hline
  Knot & \multicolumn{7}{|c||}{Simple Cubic Lattice} \\
  \hline  
   & $n$ & $\C{P}_K$ & \multicolumn{2}{|c||}{$\C{S}_K$} 
   & $\C{W}_K$ & $\C{|W|}_K$ & $\C{K}_K$ \\
   \hline
$3_1$ & 24 & 3328  & 142  & $24^{137}8^5$    
    &  \SMALL $3\frac{735}{1664}$ &  \SMALL $3\frac{735}{1664}$ 
    &  \SMALL $3 \frac{801}{1664}$ \\
$ $ & 26 & 281208  & 11721 & $24^{11713}12^8$    
    &  \SMALL $3\frac{10773}{23434}$ 
    &  \SMALL $3\frac{10773}{23434}$ 
    &  \SMALL $3 \frac{23017}{23434}$ \\
$ $ & 28 & 14398776  & 599949  & $24^{599949}$    
    &  \SMALL $3\frac{40144}{85707}$ 
    &  \SMALL $3\frac{40144}{85707}$ 
    &  \SMALL $4\frac{267364}{599949}$ \\
\hline
 \end{tabular}
\end{center}
 \caption{SC Lattice Trefoils of lengths $24$, $26$ and $26$.}
  \label{tabletrefoils}
\end{table}
%%%%%%%%%%%%%%%%%%%%%%%%%%%%%%%%%%%%%%%%%%%%%%

It is known that achiral knots have zero average writhe 
\cite{JvRSW99}, and so $\C{W}_K = 0$ if the knot type $K$ is 
achiral. For example, $\C{W}_{3_1} = 3\frac{735}{1664} \approx
3.4417\ldots$ (for right handed trefoils), and hence $3_1$ is
a chiral knot type.  This numerical estimate for $\C{W}_{3_1}$
is consistent with the results of simulations done elsewhere
\cite{JvROSTW96,JvRSW99}, and it appears that $\C{W}_{3_1}$ is
only weakly dependent on the length of the polygons.  For example,
in table \ref{tabletrefoils} the average and average absolute writhe of
polygons with knot type $3_1^+$ and lengths $24$, $26$ and $28$
are listed.  Observe that while $\C{W}_{3_1}$ and $|\C{W}|_K$
do change with increasing $n$, it is also so that the change
is small, that is, it changes from $3.44170\ldots$ for $n=24$
to $3.45971\ldots$ to $3.46848\ldots$ as $n$ increments from
$n=24$ to $n=28$.  These numerical values are close to
the estimates of average writhes made elsewhere in the literature
for polygons of significant increased length, and the average
writhe seems to cluster about the estimate $3.44\ldots$ in those
simulations \cite{JvROSTW96,BOW09,PDSAV10}.

Generally, the average and average absolute writhe increases
with crossing number in table \Ref{tableSC}.  However, in each
class of knot types of crossing number $C>3$ there are knot types
with small average absolute writhe (and thus with small average
writhe).  For example, amongst the class of knot types on eight
crossings, there are achiral knots with zero absolute writhe
($8_3$), as well as chiral knot types with average absolute
writhe small compared to the average absolute writhe of (say)
$8_1$.  For example, the average absolute writhe of $8_{18}$ is
$10/37$.  The obvious question following from this observation
is on the occurrence of such knot types:  Since there are chiral
knot types with average absolute writhe less than $1$ for knots
on $4$, $6$, $7$ and $8$ crossings in table \ref{tableSC}, would
such chiral knot types exist for all knot types on $C\geq 6$ 
crossings?

The curvature of a cubic lattice polygon is a multiple of $\pi/2$,
and hence the average curvature will similarly be a rational 
number times $2\pi$:  That is, $\LL C_K \RR = 2\pi \C{K}_K$ 
where $\LL C_K \RR$ is the average curvature of minimal length 
polygons of knot type $K$ and $\C{K}_K$ is the rational number 
displayed in the last column of table \ref{tableSC}. For example, 
the average curvature of minimal length lattice trefoils is 
$6\LH \frac{801}{832}\RH \pi \approx 6.96274\pi$.  

The variability in $\C{K}_K$ is less than that observed for
the writhe $\C{W}_K$ in classes of knot types of given crossing
number in table \ref{tableSC}.  Generally, increasing the
crossing number increases the minimal length of the knot type,
with a similar increase in the number of right angles in the polygon.
This increase is reflected in the increase of $\C{K}_K$ with
increasing $n_K$.

The ratio $\C{K}_K/n_K$ stabilizes quickly in families of knot types.
For example, for $(N,2)$-torus knots, this ratio decreases with
increasing $n_K$ as $\{0.25,0.141,0.142,0.141\}$ as $K$ increases
along $\{3_1,5_1,7_1,9_1\}$. Similar patterns can be  determined
for other families of knot types.  For example, for the twist
knots $K=\{4_1,6_1,8_1,10_1\}$, the ratio is also stable, but a little
bit lower: $\{0.145,0.136,0.131,0.136\}$.

Finally, it was observed before equation \Ref{eqn25} that the
minimal curvature of a lattice knot in the SC lattice,
$\C{C}_K$, can be defined and that $\C{C}_{0_1}=2\pi$,
$\C{C}_{3_1} = 6\pi$ and $\C{C}_{9_{47}}=9\pi$.  The average curvatures
in table \ref{tableSC} exceeds these lower bounds in general,
with equality only for the unknot:  For example, 
$\C{K}_{3_1} = 6\frac{801}{832} \pi$ and $\C{K}_{9_{47}} 
= 10\frac{94}{95} \pi$.  However, in each of these
knot types there are realizations of polygons with both minimal
length and minimal curvature.

%%%%%%%%%%%%%%%%%%%%%%%%%%%%%%%%%%%%%%%%%%%%%%%%%%%%%%%%%%%%%%%%%%
%%%%%%%%%%%%%%%%%%%%%%%%%%%%%%%%%%%%%%%%%%%%%%%%%%%%%%%%%%%%%%%%%%
\subsection{Minimal Knots in the Face Centered Cubic Lattice}

%%%%%%%%%%%%%%%%%%%%%%%%%%%%%%%
\begin{table}[t!]
\begin{center}
 \begin{tabular}{||c|| l ||}
 \hline
  $n_K$ & Prime Knot types \\
 \hline 
$3$ & $0_1$  \\ 
$15$ & $3_1$  \\
$20$ & $4_1$  \\
$22$ & $5_1$  \\
$23$ & $5_2$  \\
$27$ & $6_1$, $6_2$   \\
$28$ & $6_3$, $8_{19}$  \\
$29$ & $7_1$  \\
$30$ & $7_2$, $7_3$, $7_4$, $8_{20}$  \\
$31$ & $7_5$, $7_6$, $7_7$, $8_{21}$   \\
$32$ & $9_{42}$ \\
$34$ & $8_1$, $8_2$, $8_3$, $8_4$, $8_5$, $8_6$, 
       $8_7$, $8_8$, $8_9$, $8_{10}$  \\
$35$ & $8_{11}$, $8_{12}$, $8_{13}$, $8_{14}$, 
       $8_{15}$, $8_{16}$, $8_{17}$, $9_1$, $9_{47}$   \\
$36$ & $8_{18}$  \\
$37$ & $9_2$  \\
$40$ & $10_1$, $10_2$  \\
 \hline
 \end{tabular}
\end{center}
 \caption{Minimal Length of knot types in the FCC lattice.}
  \label{tablenKFCC}
\end{table}
%%%%%%%%%%%%%%%%%%%%%%%%%%%%

%%%%%%%%%%%%%%%%%%%%%%%%%%%%%%%%%%%%%%%%%%%%%%%%%%%%%%%%%%%%%%%%%%
\subsubsection{Minimal Length FCC Lattice Knots:}

The minimal lengths $n_K$ of prime knot types $K$ in the FCC
are displayed in table \ref{tablenKFCC}.  Prime knots types
up to eight crossings are included, together with a few knots
with nine crossings, as well as the knots $10_1$ and $10_2$. 
In general the pattern of data in table \ref{tablenKFCC} are
similar to the results in the SC lattice in table \ref{tablenKSC}.
Observe that while the knot type $6_*$ can be tied with $40$
edges in the SC lattice, in the FCC lattice $6_1$ and $6_2$ can 
be tied with fewer edges than $6_3$.  Similarly, the knot $7_1$ can 
be tied with fewer edges than other seven crossing knots in the FCC
lattice, but not in the SC lattice.   There are other similar 
minor changes in the ordering of the knot types in table 
\ref{tablenKFCC} compared to the SC lattice data in table \ref{tablenKSC}.

Similar to the argument in the SC lattice, one may define 
$Q_n$ to be the number of different knot types with 
$n_K \leq n$ in the FCC lattice. It follows that $Q_n \leq p_n$, 
so that 
\begin{equation}
\C{Q} = \limsup_{n\to\infty} Q_n^{1/n} \leq \lim_{n\to\infty} p_n^{1/n}
= \mu
\end{equation}
by equation \Ref{eqn1AA}.

By counting the number of distinct knot types 
with $n_K \leq n$ in tables \ref{tablenKFCC} one may estimate 
$\C{Q}$ by computing $Q_n^{1/n}$:  Observe that $Q_3=1$ and 
$Q_{15}=2$, this shows that $\C{Q} \approx 1.041\ldots$.   By 
increasing $n$, one finds that $Q_{35} \geq 37$, and this 
gives the estimate $\C{Q} \approx 1.10550\ldots$.  This is larger
than the estimate of $\C{Q}$ in the SC lattice, and may be some
evidence that the exponential rate of growth of $Q_n$ in the 
FCC lattice is strictly larger than in the SC lattice:
That is, $\C{Q}_{FCC} > \C{Q}_{SC}$. 

Similar to the case in the SC, the number of distinct knot types 
with $n_K = n$ is $\ol{Q}_n = Q_{n} - Q_{n-1}$, and since 
$Q_n \geq Q_{n-1}$ and $Q_n = \C{Q}^{n+o(n)}$, it follows that 
$\ol{Q}_n = \C{Q}^{n+o(n)}$.

There are several cases of (semi)-regularity amongst the 
minimal lengths of knot types in tables \ref{tablenKFCC}.
$(N,2)$-torus knots with $3 \leq N\leq 5$ (these are the knots 
$\{3_1,5_1,7_1\}$) have increases in steps of $7$ starting in 
$15$. This pattern, however, fails for the next member in this
sequence, since $n_{9_1} = 35$, an increment of $6$ from
$7_1$.  Similar observations are true of the sequence of
twist knots.  The sequence $\{4_1,6_1,8_1\}$ has 
increments of $7$ starting in $20$, but this breaks down
for $10_1$, which increments by $6$ over $n_{8_1}$.
The first three members of the sequence of twist knots 
$\{5_2,7_2,9_2\}$ similarly have increments in steps of $7$, and if 
the patterns above applies in this case as well, then this should
break down as well.  Observe that these results are different
from the results in the SC lattice.  In that case, the patterns
persisted for the knots examined, but in the FCC lattice the patterns
break down fairly quickly. 

%%%%%%%%%%%%%%%%%%%%%%%%%%%%%%%%%%%%%%%%%
\begin{table}[t!]
\begin{center}
 \begin{tabular}{||c||r|r|r|r||l|l|l||}
 \hline
  Knot & \multicolumn{7}{|c||}{Face Centered Cubic Lattice} \\
  \hline  
   & $n_K$ & $\C{P}_K$ & \multicolumn{2}{|c||}{$\C{S}_K$} 
   & $\C{W}_K$ & $\C{|W|}_K$ & $\C{K}_K $ \\
   \hline
$0_1$ & 3  & 8     & 1   & $8^1$ & \SMALL 0  & \SMALL 0 
    & \SMALL 1 \\
$3_1$ & 15& 64 & 4  & $24^{2}8^{2}$    
    &  \SMALL $3.3245203$ &  \SMALL $3.3245203$ 
    &  \SMALL $2\frac{3}{4}$ \\
$4_1$ & 20 & 2796  & 130 & $24^{106}12^{18}6^6$ 
    &  \SMALL $0$ &  \SMALL $0.0649554$ 
    &  \SMALL $3 \frac{175}{699}$ \\
$5_1$ & 22 & 96  & 4 & $24^{4}$
    &  \SMALL $6.04086733$ &  \SMALL $6.04086733$ 
    & \SMALL $3\frac{1}{2}$ \\
$5_2$ & 23 & 768 & 32 & $24^{32}$
    &  \SMALL $4.58773994$ &  \SMALL $4.58773994$ 
    & \SMALL $3\frac{3}{4}$ \\
$6_1$ & 27 & 19008  & 792 & $24^{792}$
    &  \SMALL $1.30062599$ &  \SMALL $1.30062599$ 
    & \SMALL $4\frac{449}{1188}$ \\
$6_2$ & 27 & 5040 & 210 & $24^{210}$
    &  \SMALL $2.68566969$ &  \SMALL $2.68566969$ 
    & \SMALL $4\frac{199}{630}$ \\
$6_3$ & 28 & 102720 & 4280 & $24^{4280}$
    &  \SMALL $0$ &  \SMALL $0.10145467$ 
    & \SMALL $4\frac{11519}{25680}$ \\
$7_1$ & 29 & 4080  & 170 & $24^{170}$
    &  \SMALL $8.83566369$ &  \SMALL $8.83566369$ 
    & \SMALL $4\frac{919}{1020}$ \\
$7_2$ & 30 & 4128  & 172 & $24^{172}$
    &  \SMALL $5.94373229$ &  \SMALL $5.94373229$ 
    & \SMALL $4\frac{37}{43}$ \\
$7_3$ & 30 & 960  & 40 & $24^{40}$
    &  \SMALL $7.30408669$ &  \SMALL $7.30408669$ 
    & \SMALL $4\frac{3}{5}$ \\
$7_4$ & 30 & 96 & 4 & $24^{4}$
    &  \SMALL $6.17547989$ &  \SMALL $6.17547989$ 
    & \SMALL $4\frac{5}{6}$ \\
$7_5$ & 31 & 27456 & 1144 & $24^{1144}$
    &  \SMALL $7.31767838$ &  \SMALL $7.31767838$ 
    & \SMALL $4\frac{853}{858}$ \\
$7_6$ & 31 & 4896  & 204 & $24^{204}$
    &  \SMALL $3.29853635$ &  \SMALL $3.29853635$ 
    & \SMALL $5\frac{2}{17}$ \\
$7_7$ & 32 & 1296  & 54 & $24^{54}$
    &  \SMALL $0.66279311$ &  \SMALL $0.66279311$ 
    & \SMALL $5\frac{35}{162}$ \\
$8_1$ & 34 & 447816 & 18696 & $24^{18622}12^{74}$
    &  \SMALL $2.51971823$ &  \SMALL $2.51971823$ 
    & \SMALL $5\frac{11155}{18659}$ \\
$8_2$ & 34 & 116016 & 4834 & $24^{4834}$
    &  \SMALL $5.39777682$ &  \SMALL $5.39777682$ 
    & \SMALL $5\frac{7991}{14502}$ \\
$8_3$ & 34 & 19200 & 800 & $24^{800}$
    &  \SMALL $0$ &  \SMALL $0.06471143$ 
    & \SMALL $5\frac{73}{240}$ \\
$8_4$ & 34 & 41088 & 1712 & $24^{1712}$
    &  \SMALL $1.39528958$ &  \SMALL $1.39528958$ 
    & \SMALL $5\frac{863}{1712}$ \\
$8_5$ & 34 & 2976 & 130 & $24^{118}12^{12}$
    &  \SMALL $5.40078543$ &  \SMALL $5.40078543$  
    & \SMALL $5\frac{12}{31}$ \\
$8_6$ & 34 & 9408 & 392 & $24^{392}$
    &  \SMALL $3.94736084$ &  \SMALL $3.94736084$  
    & \SMALL $5\frac{10}{21}$ \\
$8_7$ & 34 & 1258 & 52 & $24^{52}$
    &  \SMALL $2.70284845$ &  \SMALL $2.70284845$  
    & \SMALL $5\frac{21}{52}$ \\
$8_8$ & 34 & 3024 & 126 & $24^{126}$
    &  \SMALL $1.28153619$ &  \SMALL $1.28153619$  
    & \SMALL $5\frac{9}{14}$ \\
$8_9$ & 34 & 5184 & 216 & $24^{216}$
    &  \SMALL $0$ &  \SMALL $0.0808692$  
    & \SMALL $5\frac{20}{81}$ \\
$8_{10}$ & 34 & 1728 & 72 & $24^{72}$
    &  \SMALL $2.82452035$ &  \SMALL $2.82452035$  
    & \SMALL $5\frac{5}{18}$ \\
$8_{11}$ & 35 & 298128 & 12422 & $24^{12422}$
    &  \SMALL $3.97223690$ &  \SMALL $3.97223690$  
    & \SMALL $5\frac{11713}{18633}$ \\
$8_{12}$ & 35 & 16416 & 684 & $24^{684}$
    &  \SMALL $0.13164234$ &  \SMALL $0.13164234$  
    & \SMALL $5\frac{173}{229}$ \\
$8_{13}$ & 35 & 274320 & 11430 & $24^{11430}$
    &  \SMALL $1.30541189$ &  \SMALL $1.30541189$  
    & \SMALL $5\frac{9256}{17145}$ \\
$8_{14}$ & 35 & 27360 & 1140 & $24^{1140}$
    &  \SMALL $4.00297606$ &  \SMALL $4.00297606$  
    & \SMALL $5\frac{1109}{1710}$ \\
$8_{15}$ & 35 &36432 & 1518 & $24^{1518}$
    &  \SMALL $7.98074463$  &  \SMALL $7.98074463$ 
    & \SMALL $5\frac{215}{414}$ \\
$8_{16}$ & 35 & 15552 & 648 & $24^{648}$
    &  \SMALL $2.66666668$ &  \SMALL $2.66666668$ 
    & \SMALL $5\frac{17}{36}$ \\
$8_{17}$ & 35 & 5184 & 216 & $24^{216}$
    &  \SMALL $0$ &  \SMALL $0.08782937$ 
    & \SMALL $5\frac{35}{108}$ \\
$8_{18}$ & 36 & 41196 & 1776 & $24^{1662}12^{104}6^{10}$
    &  \SMALL $0$ &  \SMALL $0.12891984$ 
    & \SMALL $5\frac{1817}{3433}$ \\
$8_{19}$ & 28 & 276 & 12 & $24^{11}12^{1}$
    &  \SMALL $8.45506005$ &  \SMALL $8.45506005$ 
    & \SMALL $4$ \\
$8_{20}$ & 30 & 74088 & 3087 & $24^{3087}$
    &  \SMALL $2.04596806$ &  \SMALL $2.04596806$ 
    & \SMALL $4\frac{3137}{6174}$ \\
$8_{21}$ & 31 & 17856 & 744 & $24^{744}$
    &  \SMALL $4.66448881$ &  \SMALL $4.66448881$ 
    & \SMALL $4\frac{2039}{2232}$ \\
$9_1$ & 35 & 192 & 8 & $24^{8}$
    &  \SMALL $11.58173466$ &  \SMALL $11.58173466$
    & \SMALL $5\frac{3}{4}$ \\
$9_2$ & 37 & 229824 & 9576 & $24^{9576}$
    &  \SMALL $7.13091283$ &  \SMALL $7.13091283$
    & \SMALL $6\frac{2}{21}$ \\
$9_{42}$ & 32 & 96 & 4 & $24^{4}$
    &  \SMALL $1.02043366$ &  \SMALL $1.02043366$
    & \SMALL $4\frac{5}{12}$ \\
$9_{47}$ & 35 & 3072 & 128 & $24^{128}$
    &  \SMALL $2.62349866$ &  \SMALL $2.62349866$
    & \SMALL $5\frac{19}{48}$ \\
$10_1$ & 40 & 77688 & 3246 & $24^{3228}12^{18}$
    &  \SMALL $3.74646905$ &  \SMALL $3.74646905$
    & \SMALL $6\frac{11237}{19422}$ \\
$10_2$ & 40 & 8928 & 372 & $24^{372}$
    &  \SMALL $8.14499622$ &  \SMALL $8.14499622$
    & \SMALL $6\frac{671}{1116}$ \\
\hline
 \end{tabular}
\end{center}
 \caption{Data on prime knot types in the FCC Lattice.}
  \label{tableFCC}
\end{table}
%%%%%%%%%%%%%%%%%%%%%%%%%%%%%%%%%%%%%%%%%%%%%%%%%%%%%%%%%%

%%%%%%%%%%%%%%%%%%%%%%%%%%%%%%%%%%%%%%%%%%%%%%%%%%%%%%%%%%%%%%%%%%
\subsubsection{Entropy of minimal lattice knots in the FCC Lattice:}

Data on entropy on minimal length polygons were 
collected of FCC lattice polygons with prime knot types up to 
eight crossings, and also knot the knots $9_1$, $9_1$, $9_{42}$,
$9_{47}$, $10_1$ and $10_2$.  The results are displayed in
table \ref{tableFCC}. The minimal length of a knot type $K$ is 
denoted by $n_K$, and $\C{P}_K = c_{n_K}(K)$ is the total number 
of minimal length FCC lattice knots of length $n_K$.  
For example, there are $64$ minimal length trefoils 
(of both chiralities) of length $n_{3_1}=15$ in the FCC lattice.
Since $3_1$ is chiral, $\C{P}_{3_1^+} = 64/2 = 32$.  

Each set of minimal length lattice knots are divided into symmetry
classes under action of the symmetry group of rotations and reflections
in the FCC lattice.  For example, the unknot has minimal length
$3$ and it is a member of a symmetry class of $8$ FCC lattice polygons
of minimal length which are equivalent under action of the 
symmetry elements of the octahedral group.

The $64$ minimal length FCC lattice trefoils are similarly divided
into $4$ symmetry classes, of which $2$ classes have $24$ members
and $2$ classes have $8$ members each (which are symmetric under
action of some of the group elements).  This partitioning into 
symmetry classes are denoted by $8^1$ for the unknot, and $24^{2}
8^2$ for the trefoil (of both chiralities).

Similar to the case for the SC lattice, the reliability of the
data in table \ref{tableFCC} decreases with increasing values
of $\C{P}_K$.  We are very certain of our data if 
$\C{P}_K \lesssim 1000$, reasonable certain if 
$1000 \lesssim \C{P}_K \lesssim 10000$, less certain if 
$10000\lesssim \C{P}_K \lesssim 100000$, and 
we consider the stated value of $\C{P}_K$ to be only a lower
bound if $\C{P}_K \gtrsim 100000$ in table \ref{tableFCC}.

The entropy per unit length of minimal polygons of knot type 
$K$ is similarly defined in this lattice in equation \Ref{eqn31}.
The unknot has relative large entropy: 
$\C{E}_{0_1} = (\log 8)/3 = 0.693147\ldots$.

The entropy per unit length of the (right-handed) trefoil
is $\C{E}_{3_1^+} = (\log  32)/15 = 0.2310\ldots$, which is smaller
than the entropy of this knot type in the SC.  This implies that
there are fewer conformations per edge, and the knot may be 
considered to be more tightly embedded. 

The entropy per unit length of the (achiral) knot $4_1$ 
is $\C{E}_{4_1} = (\log 2796)/20 = 0.3968\ldots$, and is 
less than the trefoil (however, in the SC lattice 
$\C{E}(3_1^+) > \C{E}(4_1)$).  For five crossing knots 
one finds that $\C{E}_{5_1^+} = 0.1760\ldots$ while 
$\C{E}_{5_2^+} = 0.2587\ldots$; these are related similarly
to the results in the SC lattice.

The entropy per unit length in the family of  $(N,2)$-torus knots
$\{3_1^+,5_1^+,7_1^+,9_1^+\}$ changes as  
$\{0.2310,0.1760,0.2628,0.1304\}$ to four digits accuracy.
These results do not show the regularity observed in the SC:  While
the results for $\{3_1^+,5_1^+,9_1^+\}$ decreases in sequence,
the result for $7_1^+$ seems to be unrelated.

The family of twist knots $\{3_1^+,5_2^+,7_2^+,9_2^+\}$
gives $\{0.2310,0.2587,0.2544,0.3149\}$, again to four digits,
and this case the knot $9_2^+$ seems to have a value higher than
expected.  Similar observations can be made for the families
$\{6_1^+,8_1^+,10_1^+\}$ ($\{0.3392,0.3623,0.2642\}$), 
and $\{6_2^+,8_2^+,10_2^+\}$ ($\{0.2901,0.3226,0.2101\}$).  
Further extensions of the estimates of $\C{P}_K$ for more 
complicated knots would be necessary to determine if any of
these sequences approach a limiting value.

Finally, there are some knots with very low entropy per unit length.
These include $7_4^+$ ($0.1290$), $9_1^+$ ($0.1106$), 
and $9_{42}^+$ ($0.1210$). These knots are tightly embedded in the 
FCC lattice in their minimal conformations, with very little entropy 
per edge available.

The distribution of minimal knotted polygons in symmetry classes
in table \ref{tableFCC} shows that most minimal knotted polygons
are not symmetric with respect to elements of the octahedral 
group, and thus fall into classes of $24$ distinct polygons.
Classes with fewer elements, (for example $12$ or $8$), 
contains symmetric embeddings of the embedded polygons.  Such
symmetric embeddings are the exception rather than the rule in
table \ref{tableFCC}:  This is similar to the observations made
in the SC lattice. 

%%%%%%%%%%%%%%%%%%%%%%%%%%%%%%%%%%%%%%%%%%%%%%%%%%%%%%%%%
\begin{table}[t!]
\begin{center}
 \begin{tabular}{||c||r|r|r|r||l|l|l||}
 \hline
  Knot & \multicolumn{7}{|c||}{Face Centred Cubic Lattice} \\
  \hline  
   & $n$ & $\C{P}_K$ & \multicolumn{2}{|c||}{$\C{S}_K$} 
   & $\C{W}_K$ & $\C{|W|}_K$ & $\C{K}_K$ \\
   \hline
$3_1$ & 15& 64 & 4  & $24^{2}8^{2}$    
    &  \SMALL $3.3245203$ &  \SMALL $3.3245203$ 
    &  \SMALL $2\frac{3}{4}$ \\
$ $ & 16& 3672 & 153  & $24^{153}$    
    &  \SMALL $3.34714432$ &  \SMALL $3.34714432$ 
    &  \SMALL $2\frac{404}{459}$ \\ 
$ $ & 17& 104376 & 4349  & $24^{4349}$    
    &  \SMALL $3.36103672$ &  \SMALL $3.36103672$ 
    &  \SMALL $3\frac{853}{13047}$ \\
\hline
 \end{tabular}
\end{center}
 \caption{Data on trefoils of lengths $15$, $16$ and $17$
 in the FCC lattice.}
  \label{tabletrefoilsfcc}
\end{table}
%%%%%%%%%%%%%%%%%%%%%%%%%%%%%%%%%%%%%%%%%%%%%%%%%%%%%%%%%%%%%%%%%%%%

%%%%%%%%%%%%%%%%%%%%%%%%%%%%%%%%%%%%%%%%%%%%%%%%%%%%%%%%%%%%%
\subsubsection{The Lattice Writhe and Curvature in the FCC Lattice:}

The average writhe $\C{W}_K$, the average absolute writhe $|\C{W}|_K$
and the average curvature $\C{K}_K$ (in units of $2\pi$)
of minimal length FCC lattice polygons are displayed in table
\ref{tableFCC}.  The results for the average writhe are given 
in floating point numbers since these are irrational numbers 
in the FCC lattice, as seen for example from equation
\Ref{eqn21}.  

The lattice curvature of a given FCC lattice polygon,
on the other hand, is a multiple of $\pi/3$, and thus $2\pi \C{K}_K$,
where $\C{K}_K$ is average curvature, is a rational number.  In
table \ref{tableFCC} the average curvature $\C{K}_K$ is given
in units of $2\pi$, so that the exact values of this average
quantity can be given as a rational number.  For example, one
infers from table \ref{tableFCC} that the average curvature of 
the unknot is $2\pi$, while the average curvature of $3_1$ is
$2\frac{3}{4} (2\pi) = 5\frac{1}{2}\pi$.

Similar to the results in the SC lattice, the absolute average and
average absolute writhes in table \ref{tableSC} are equal, except
for achiral knots.  This pattern may break down eventually, but
persists for the knots we considered. In the case of achiral knots
one has, as for the SC lattice, $\C{W}_K = 0$ while $|\C{W}|_K > 0$.
Observe that the average absolute writhe of $8_3$ is positive in 
the FCC, but it is zero in the SC.

The average writhe at minimal length of $3_1^+$ is $3.3245\ldots$ 
in the FCC, while it is slightly larger in the SC, namely 
$3.4417$.  Increasing the value of $n$ from $15$ to $16$ and $17$ 
in the FCC lattice and measuring the average writhe gives the 
results in table \ref{tabletrefoilsfcc}, which shows that the 
average writhe increases slowly with $n$. However, the average 
writhe remains, as in the SC lattice, quite insensitive to $n$.

Generally, the average and average absolute writhe increases
with crossing number in table \ref{tableFCC}.  However, in each
class of knot types of crossing number $C>3$ there are knot types
with small average absolute writhe (and thus with small average
writhe).  For example, amongst the class of knot types on eight
crossings, there are achiral knots with small absolute writhe
($8_3$), as well as chiral knot types with average absolute
writhe small compared to the average absolute writhe of (say)
$8_1$.  For example, the average absolute writhe of $8_9$, $8_{17}$ 
and $8_{18}$ are small compared to  other eight crossing knots
(except $8_3$).

While the average writhe is known not to be rational in the
FCC, it is nevertheless interesting to observe that the
average writhe of $8_{16}$ is almost exactly $8/3$ 
(it is approximately $41472.00002289/15552 = 
8.0000000044/3$). Similarly, the average writhe of the figure 
eight knot is very close to $13/200$ 
(it is approximately $12.99108/200$).

The average curvature $\C{K}_K$ tends to increase consistently 
with $n_K$ and with crossing number of $K$. The ratio 
$\C{K}_K/n_K$ stabilizes quickly in families of knot types. 
For example, for $(N,2)$-torus knots, 
this ratio decreases with increasing $n_K$ as 
$\{0.183,0.159,0.169,0.164\}$ as $K$ increases
along $\{3_1,5_1,7_1,9_1\}$. These estimates are slightly larger
than the similar estimates in the SC lattice.  
Similar patterns can be  determined for other families of knot 
types.  For example, for the twist knots $K=\{4_1,6_1,8_1,10_1\}$, 
the ratio is also stable and close in value to the twist
knot results: $\{0.163,0.162,0.165,0.164\}$.

Finally, the average curvature of the trefoil in the FCC is
$5\frac{1}{2}\pi$ and this is less than the lower bound $6\pi$ 
of the minimal curvature of a trefoil in the SC lattice \cite{JvRP99}.
The minimal curvature of $9_{47}$ at minimal length in the SC lattice 
is $9\pi$ \cite{JvRP99}, but in the FCC lattice our data show 
no FCC polygons of knot type $9_{47}$ and minimal length $n=35$
has curvature less than $10\frac{1}{6}\pi$.  In other words,
there is no realisation of a polygon of knot type $9_{47}$ in the 
FCC at minimal length $n_K=35$ with minimal curvature 
$9\pi$.  The average curvature of minimal length FCC lattice 
knots of type $9_{47}$ is still larger than the these 
lower bounds, namely $10\frac{19}{24} \pi$.

%%%%%%%%%%%%%%%%%%%%%%%%%%%%%%%%%%%%%%%%%%%%%%%
\subsection{Minimal Knots in the Body Centered Cubic Lattice}

%%%%%%%%%%%%%%%%%%%%%%%%%%%%%%%

\begin{table}[t!]
\begin{center}
 \begin{tabular}{||c|| l ||}
 \hline
  $n_K$ & Prime Knot types  \\
 \hline 
$4$ & $0_1$ \\ 
$18$ & $3_1$ \\
$20$ & $4_1$ \\
$26$ & $5_1$, $5_2$ \\
$28$ & $6_1$ \\
$30$ & $6_2$, $6_2$ \\
$32$ & $7_1$, $7_2$, $7_6$, $7_7$, $8_{19}$  \\
$34$ & $7_3$, $7_4$, $7_5$, $8_{20}$, $8_{21}$  \\
$36$ & $8_1$, $8_3$, $8_{12}$ \\
$38$ & $8_2$, $8_4$, $8_5$, $8_6$, $8_7$, $8_8$, $8_9$, $8_{10}$ \\
$38$ & $8_{11}$, $8_{13}$, $8_{14}$, $8_{15}$, $8_{16}$, $8_{17}$  \\
$40$ & $8_{18}$, $9_1$, $9_2$ \\
$42$ & $10_{1}$ \\
$44$ & $10_{2}$ \\
 \hline
 \end{tabular}
\end{center}
 \caption{Minimal Length of Knots types in the BCC lattice.}
  \label{tablenKBCC}
\end{table}
%%%%%%%%%%%%%%%%%%%%%%%%%%%%

%%%%%%%%%%%%%%%%%%%%%%%%%%%%%%%%%%%%%%%%%%%%%%%%%%%%%%%%%%%%%%%%%%%%%
%%%%%%%%%%%%%%%%%%%%%%%%%%%%%%%%%%%%%%%%%%%%%%%%%%%%%%%%%%%%%%%%%%%%%
\subsubsection{Minimal Length BCC Lattice Knots:}

The minimal lengths $n_K$ of prime knot types $K$ in the BCC
are displayed in table \ref{tablenKBCC}.  We again included
prime knot types up to eight crossings, together with a few knots
with nine crossings, as well as the knots $10_1$ and $10_2$. 

In general the pattern of data in table \ref{tablenKBCC} is
similar to the results in the SC and FCC lattices in tables
\ref{tablenKSC} and \ref{tablenKBCC}.  The spectrum of knots
corresponds well up to five crossings, but again at six
crossings some differences appear.  For example, in the BCC lattice
one observes that $n_{6_1} < n_{6_2}$ and $n_{6_1} < n_{6_3}$,
in contrast with the patterns observed in the SC and FCC lattices.

The rate of increase in the number of knot types of minimal 
length $n_K\leq n$ in the BCC lattice may be analysed in the
same way as in the SC or FCC lattice.  Similar to the argument in 
the SC lattice, one may define 
$Q_n$ to be the number of different knot types with 
$n_K \leq n$ in the FCC lattice. It follows that $Q_n \leq p_n$, so that 
\begin{equation}
\C{Q} = \limsup_{n\to\infty} Q_n^{1/n} \leq \lim_{n\to\infty} p_n^{1/n}
= \mu
\end{equation}
by equation \Ref{eqn1AA}.

By counting the number of distinct knot types 
with $n_K \leq n$ in table \ref{tablenKBCC} one may estimate
$\C{Q}$:  Observe that $Q_4=1$ and $Q_{18}=2$, 
this shows that $\C{Q} \approx 2^{1/16} = 1.035\ldots$.  
By increasing $n$ while counting knot types to estimate $Q_n$, 
one finds that $Q_{38} \geq 35$, and this gives the lower bound 
$\C{Q} \approx 35^{1/40} = 1.0929\ldots$.  This is larger
than the lower bound on $\C{Q}$ in the SC lattice, and may
again be taken as evidence that $Q_n$ is exponentially small in the 
SC lattice when compared to the BCC lattice.  That is
$\C{Q}_{SC} < \C{Q}_{BCC}$.  

Similar to the case in the SC lattice, the number of distinct knot 
types with $n_K = n$ is $\ol{Q}_n = Q_{n} - Q_{n-1}$, and since 
$Q_n \geq Q_{n-1}$ and $Q_n = \C{Q}^{n+o(n)}$, it follows that 
$\ol{Q}_n = \C{Q}^{n+o(n)}$ for even values of $n$ (note that
$\ol{Q}_n = 0$ of $n$ is odd, since the BCC is a bipartite lattice).

There are several cases of (semi)-regularity amongst the 
minimal lengths of knot types in tables \ref{tablenKFCC}.
$(N,2)$-torus knots (these are the knots 
$\{3_1,5_1,7_1,9_1\}$) increase in steps of $6$ or $8$ starting in 
$18$. The increments are $\{8,6,8\}$ in this particular case,
and there are no indications that this will be repeating,
or whether it will persist at all. Similar observations are true 
of the sequence of twist knots.  The sequence $\{4_1,6_1,8_1,10_1\}$
seems to have increments of $8$ starting in $20$, but this breaks 
down for $10_1$, which increments by $6$ over $8_1$.  Similar
observations can be made for the sequence of twist knots 
$\{5_2,7_2,9_2\}$. 

%%%%%%%%%%%%%%%%%%%%%%%%%%%%%%%%%%%%%%%%%
\begin{table}[t!]
\begin{center}
 \begin{tabular}{||c||r|r|r|r||l|l|l||}
 \hline
  Knot & \multicolumn{7}{|c||}{Body Centered Cubic Lattice} \\
  \hline  
   & $n_K$ & $\C{P}_K$ & \multicolumn{2}{|c||}{$\C{S}_K$} 
   & $\C{W}_K$ & $\C{|W|}_K$ & $\C{B}_K,\C{K}_K$ \\
   \hline
$0_1$ & 4  & 12     & 2   & $6^2$ & \SMALL 0  & \SMALL 0 
    & \SMALL $-2$, $\frac{3}{2}$ \\
$3_1$ & 18 & 1584 & 66  & $24^{66}$    
    &  \SMALL $3\frac{40}{99}$ &  \SMALL $3\frac{40}{99}$ 
    &  \SMALL $11\frac{4}{33}$, $\frac{21}{22}$ \\
$4_1$ & 20& 12 & 2  & $6^2$    
    &  \SMALL $0$ &  \SMALL $0$ 
    &  \SMALL $16$, $0$ \\
$5_1$ & 26 & 14832  & 618 & $24^{618}$
    &  \SMALL $6\frac{83}{1854}$ &  \SMALL $6\frac{83}{1854}$ 
    & \SMALL $19\frac{38}{103}$, $\frac{177}{206}$ \\
$5_2$ & 26 & 4872 & 203 & $24^{203}$
    &  \SMALL $4\frac{129}{203}$ &  \SMALL $4\frac{129}{203}$ 
    & \SMALL $17\frac{23}{203}$, $\frac{164}{203}$ \\
$6_1$ & 28 & 72  & 4 & $24^{2}12^{2}$
    &  \SMALL $1\frac{1}{3}$ &  \SMALL $1\frac{1}{3}$ 
    & \SMALL $24$, $0$ \\
$6_2$ & 30 & 8256 & 344 & $24^{344}$
    &  \SMALL $2\frac{30}{43}$ &  \SMALL $2\frac{30}{43}$ 
    & \SMALL $20\frac{21}{43}$, $\frac{35}{43}$ \\
$6_3$ & 30 & 3312 & 138 & $24^{138}$
    &  \SMALL $0$ &  \SMALL $\frac{4}{69}$ 
    & \SMALL $19\frac{35}{69}$, $\frac{56}{69}$ \\
$7_1$ & 32 & 1464  & 61 & $24^{61}$
    &  \SMALL $9$ &  \SMALL $9$ 
    & \SMALL $24\frac{38}{61}$, $1$ \\
$7_2$ & 32 & 24  & 1 & $24^{1}$
    &  \SMALL $6$ &  \SMALL $6$ 
    & \SMALL $28$, $0$ \\
$7_3$ & 34 & 22488 & 937 & $24^{937}$
    &  \SMALL $7\frac{919}{2811}$ &  \SMALL $7\frac{919}{2811}$ 
    & \SMALL $\frac{25}{55}{937}$, $\frac{745}{937}$ \\
$7_4$ & 34 & 11208 & 468 & $24^{466}12^2$
    &  \SMALL $5\frac{464}{467}$ &  \SMALL $5\frac{464}{467}$ 
    & \SMALL $24\frac{394}{467}$, $\frac{340}{467}$ \\
$7_5$ & 34 & 8784 & 366 & $24^{366}$
    &  \SMALL $7\frac{196}{549}$ &  \SMALL $7\frac{196}{549}$ 
    & \SMALL $22\frac{24}{61}$, $\frac{47}{61}$ \\
$7_6$ & 32 & 48  & 2 & $24^{2}$
    &  \SMALL $3\frac{1}{3}$ &  \SMALL $3\frac{1}{3}$ 
    & \SMALL $26$, $0$ \\
$7_7$ & 32 & 24  & 1 & $24^{1}$
    &  \SMALL $\frac{2}{3}$ &  \SMALL $\frac{2}{3}$ 
    & \SMALL $24$, $0$ \\
$8_1$ & 36 & 744 & 32 & $24^{30}12^{2}$
    &  \SMALL $2\frac{2}{3}$ &  \SMALL $2\frac{2}{3}$ 
    & \SMALL $32$, $0$ \\
$8_2$ & 38 & 118080 & 4920 & $24^{4920}$
    &  \SMALL $5\frac{782}{1845}$ &  \SMALL $5\frac{782}{1845}$ 
    & \SMALL $28\frac{153}{410}$, $\frac{431}{492}$ \\
$8_3$ & 36 & 108 & 6 & $24^{4}6^2$
    &  \SMALL $0$ &  \SMALL $0$ 
    & \SMALL $32$, $0$ \\
$8_4$ & 38 & 93984 & 3916 & $24^{3916}$
    &  \SMALL $1\frac{4715}{11748}$ &  \SMALL $1\frac{4715}{11748}$ 
    & \SMALL $27\frac{955}{1958}$, $\frac{3849}{3916}$ \\
$8_5$ & 38 & 7392 & 318 & $24^{298}12^{20}$
    &  \SMALL $5\frac{331}{924}$ &  \SMALL $5\frac{331}{924}$  
    & \SMALL $29\frac{11}{14}$, $\frac{195}{308}$ \\
$8_6$ & 38 & 9024 & 376 & $24^{376}$
    &  \SMALL $4\frac{1}{282}$ &  \SMALL $4\frac{1}{282}$  
    & \SMALL $28\frac{87}{94}$, $\frac{117}{188}$ \\
$8_7$ & 38 & 47856 & 1994 & $24^{1994}$
    &  \SMALL $2\frac{2035}{2991}$ &  \SMALL $2\frac{2035}{2991}$  
    & \SMALL $27\frac{59}{997}$, $\frac{784}{997}$ \\
$8_8$ & 38 & 34656 & 1444 & $24^{1444}$
    &  \SMALL $1\frac{112}{361}$ &  \SMALL $1\frac{112}{361}$  
    & \SMALL $26\frac{177}{361}$, $\frac{280}{361}$ \\
$8_9$ & 38 & 5712 & 238 & $24^{238}$
    &  \SMALL $0$ &  \SMALL $\frac{1}{14}$  
    & \SMALL $26\frac{55}{119}$, $\frac{185}{238}$ \\
$8_{10}$ & 38 & 11088 & 462 & $24^{462}$
    &  \SMALL $2\frac{313}{462}$ &  \SMALL $2\frac{313}{462}$  
    & \SMALL $25\frac{6}{7}$, $\frac{125}{154}$ \\
$8_{11}$ & 38 & 15888 & 662 & $24^{662}$
    &  \SMALL $4 \frac{49}{1986}$ &  \SMALL $4\frac{49}{1986}$  
    & \SMALL $27\frac{198}{331}$, $\frac{425}{662}$ \\
$8_{12}$ & 36 & 12 & 2 & $6^{2}$
    &  \SMALL $0$ &  \SMALL $0$  
    & \SMALL $24$, $0$ \\
$8_{13}$ & 38 & 17616 & 734 & $24^{734}$
    &  \SMALL $1\frac{241}{734}$ &  \SMALL $1\frac{241}{734}$  
    & \SMALL $25\frac{180}{367}$, $\frac{561}{734}$ \\
$8_{14}$ & 38 & 16944 & 706 & $24^{706}$
    &  \SMALL $4\frac{1}{353}$ &  \SMALL $4\frac{1}{353}$  
    & \SMALL $25\frac{205}{353}$, $\frac{253}{353}$ \\
$8_{15}$ & 38 & 4272 & 180 & $24^{176}12^4$
    &  \SMALL $8\frac{1}{89}$  &  \SMALL $8\frac{1}{89}$ 
    & \SMALL $24\frac{52}{89}$, $\frac{71}{89}$ \\
$8_{16}$ & 38 & 1056 & 44 & $24^{44}$
    &  \SMALL $2\frac{29}{44}$ &  \SMALL $2\frac{29}{44}$ 
    & \SMALL $27\frac{15}{22}$, $\frac{23}{44}$ \\
$8_{17}$ & 38 & 912 & 38 & $24^{38}$
    &  \SMALL $0$ &  \SMALL $\frac{7}{114}$ 
    & \SMALL $24\frac{9}{19}$, $\frac{31}{38}$ \\
$8_{18}$ & 40 & 8820 & 384 & $24^{354}12^{24}6^{6}$
    &  \SMALL $0$ &  \SMALL $\frac{94}{735}$ 
    & \SMALL $24\frac{116}{735}$, $1\frac{81}{245}$ \\
$8_{19}$ & 32 & 1110 & 48 & $24^{45}12^{2}6^1$
    &  \SMALL $8\frac{102}{185}$ &  \SMALL $8\frac{102}{185}$ 
    & \SMALL $23\frac{29}{185}$, $\frac{64}{185}$ \\
$8_{20}$ & 34 & 117096 & 4879 & $24^{4879}$
    &  \SMALL $2\frac{372}{4879}$ &  \SMALL $2\frac{372}{4879}$ 
    & \SMALL $19\frac{4575}{4879}$, $1\frac{614}{4879}$ \\
$8_{21}$ & 34 & 696 & 30 & $24^{28}12^2$
    &  \SMALL $4\frac{43}{87}$ &  \SMALL $4\frac{43}{87}$ 
    & \SMALL $23\frac{23}{29}$, $\frac{14}{29}$ \\
$9_1$ & 40 & 80928 & 3372 & $24^{3372}$
    &  \SMALL $11\frac{14113}{20232}$ &  \SMALL $11\frac{14113}{20232}$
    & \SMALL $32\frac{3287}{3372}$, $1\frac{2719}{3372}$ \\
$9_2$ & 40 & 13824 & 576 & $24^{576}$
    &  \SMALL $7\frac{1}{192}$ &  \SMALL $7\frac{1}{192}$
    & \SMALL $29\frac{211}{288}$, $1\frac{77}{96}$ \\
$9_{42}$ & 36 & 2736 & 114 & $24^{114}$
    &  \SMALL $1\frac{29}{171}$ 
    &  \SMALL $1\frac{29}{171}$
    & \SMALL $24\frac{5}{57}$, $\frac{56}{57}$ \\
$9_{47}$ & 40 & 68208 & 2842 & $24^{2842}$
    &  \SMALL $2\frac{2033}{2842}$ 
    &  \SMALL $2\frac{2033}{2842}$ 
    & \SMALL $22\frac{1}{49}$, $3\frac{296}{1421}$ \\
$10_1$ & 42 & 288 & 12 & $24^{12}$
    &  \SMALL $3\frac{2}{3}$ &  \SMALL $3\frac{2}{3}$
    & \SMALL $38$, $0$ \\
$10_2$ & 44 & 9816 & 409 & $24^{409}$
    &  \SMALL $8\frac{136}{409}$ &  \SMALL $8\frac{136}{409}$
    & \SMALL $34\frac{150}{409}$, $1\frac{299}{409}$ \\
\hline
 \end{tabular}
\end{center}
 \caption{Data on knots in the BCC lattice.}
  \label{tableBCC}
\end{table}
%%%%%%%%%%%%%%%%%%%%%%%%%%%%%%%%%%%%%%%%%%%%%%%%%%%%%%%%%%

%%%%%%%%%%%%%%%%%%%%%%%%%%%%%%%%%%%%%%%%%%%%%%%%%%%%%%%%%%%%%%%%%%
\subsubsection{Entropy of minimal lattice knots in the BCC lattice:}

Data on entropy of minimal length polygons in the BCC lattice
are displayed in table \ref{tableBCC}. The minimal length of a 
knot type $K$ is denoted by $n_K$, and $\C{P}_K = c_{n_K}(K)$ is 
the total number of minimal length BCC lattice knots of length 
$n_K$.  For example, there are $1584$ minimal length trefoils 
(of both chiralities) of length $n_{3_1}=18$ in the FCC lattice.
Since $3_1$ is chiral, $\C{P}_{3_1^+} = 1584/2 = 792$.  

Each set of minimal length lattice knots are divided into symmetry
classes under the symmetry group of rotations and reflections
in the BCC lattice.  For example, the unknot has minimal length
$4$ and there are two symmetry classes, each consisting of $6$ 
BCC lattice polygons of length $4$ which are equivalent 
under action of the elements of the octahedral group.

The $1584$ minimal length BCC lattice trefoils are similarly divided
into $66$ symmetry classes, each with $24$ members.  This 
partitioning into symmetry classes are denoted by $24^{66}$ 
($66$ equivalence classes of minimal length $18$ and with
$24$ members).  Similarly, the symmetry classes of the unknot
are denoted $6^2$, namely $2$ symmetry classes of minimal length
unknotted polygons, each class with $6$ members equivalent under
reflections and rotations of the octahedral group.

Similar to the case for the SC and FCC lattices, the reliability of 
the data in table \ref{tableBCC} decreases with increasing values
of $\C{P}_K$.  We are very certain of our data if 
$\C{P}_K \lesssim 1000$, reasonable certain if 
$1000 \lesssim \C{P}_K \lesssim 10000$, less certain if 
$10000\lesssim \C{P}_K \lesssim 100000$, and 
we consider the stated value of $\C{P}_K$ to be only a lower
bound if $\C{P}_K \gtrsim 100000$.

The entropy per unit length of minimal polygons of knot type 
$K$ is similarly defined in this lattice as in equation \Ref{eqn31}.
The unknot has relative large entropy $\C{E}_{0_1}
= (\log 12)/4 = 0.621226\ldots$, compared to the entropy of the 
minimal length unknot in the SC lattice.

The entropy per unit length of the (right-handed) trefoil
is $\C{E}_{3_1^+} = (\log  792)/18 = 0.37080\ldots$, which is smaller
than the entropy of this knot type in the SC lattice.  This implies 
that there are fewer conformations per unit length, and the knot 
may be considered to be more tightly embedded.  

The entropy per unit length of the (achiral) knot $4_1$ is 
$\C{E}_{4_1} = (\log 12)/20 = 0.12424\ldots$, and is very small
compared to the values obtained in the SC and FCC lattices.
In contrast with the FCC, the relationship between the knot
types $3_1^+$ and $4_1$ in the BCC lattice is similar to the 
relationship obtained in the SC lattice, $\C{E}(3_1^+) > \C{E}(4_1)$.  
Five crossing knots in the BCC lattice have relative large
entropies.  One finds that $\C{E}_{5_1^+} = 0.34274\ldots$
while $\C{E}_{5_2^+} = 0.29992\ldots$.

The entropy per unit length in the family of  $(N,2)$-torus knots
$\{3_1^+,5_1^+,7_1^+,9_1^+\}$ changes as  
$\{0.3708,0.3427,0.2061,0.2652\}$ to four digits accuracy.
These results do not show the regularity observed in the SC lattice:
While the results for $\{3_1^+,5_1^+,7_1^+\}$ decreases in sequence,
the result for $9_1^+$ seems to buck this trend.

The family of twist knots $\{3_1^+,5_2^+,7_2^+,9_2^+\}$
gives $\{0.3708,0.3000,0.0777,0.2210\}$, again to four digits,
and this case the knot $7_2^+$ seems to have a value lower than
expected.  Similar observations can be made for the families
$\{6_1^+,8_1^+,10_1^+\}$ ($\{0.1280,0.1644,0.1183\}$), 
and $\{6_2^+,8_2^+,10_2^+\}$ ($\{0.2775,0.2891,0.1932\}$).  
Further extensions of the estimates of $\C{P}_K$ for more 
complicated knots would be necessary to determine if any of
these sequences approach a limiting value.

Finally, there are some knots with very low entropy per unit length.
These include $4_1$ ($0.1242$), $6_1^+$ ($0.1280$), 
$7_2^+$ ($0.0777$), $7_6^+$ ($0.0993$), $7_7^+$ ($0.0777$), and
$8_{12}^+$ ($0.0498$). These knots are tightly embedded in the 
BCC lattice in their minimal conformations, with very little entropy 
per edge available.

The distribution of minimal length knotted polygons in symmetry classes
in table \ref{tableFCC} shows that most minimal knotted polygons
are not symmetric with respect to elements of the octahedral 
group, and thus fall into classes of $24$ distinct polygons.
Classes with fewer elements, (for example $12$ or $8$), 
contains symmetric embeddings of the embedded polygons.  Such
symmetric embeddings are the exception rather than the rule in
table \ref{tableBCC}:  This is similar to the observations made
in the SC and FCC lattices. 

%%%%%%%%%%%%%%%%%%%%%%%%%%%%%%%%%%%%%%%%%%%%%%%%%%%%%%%%%
\begin{table}[t!]
\begin{center}
 \begin{tabular}{||c||r|r|r|r||l|l|l||}
 \hline
  Knot & \multicolumn{7}{|c||}{Body Centered Cubic Lattice} \\
  \hline  
   & $n$ & $\C{P}_K$ & \multicolumn{2}{|c||}{$\C{S}_K$}
   & $\C{W}_K$ & $\C{|W|}_K$ & $\C{B}_K,\C{K}_K$ \\
   \hline
$3_1$ & 18& 1583 & 66  & $24^{66}$    
    &  \SMALL $3\frac{40}{99}$ &  \SMALL $3\frac{40}{99}$ 
    &  \SMALL $11\frac{4}{33}$, $\frac{21}{22}$ \\
$ $ & 20& 236928 & 9879  & $24^{9865}12^{14}$    
    &  \SMALL $3\frac{22457}{59232}$ &  \SMALL $3\frac{22457}{59232}$ 
    &  \SMALL $9\frac{8369}{9872}$, $1\frac{12133}{19744}$ \\ 
$ $ & 22& 21116472 & 879864  & $24^{879842}12^{22}$    
    &  \SMALL $3\frac{1050094}{2639559}$   
    &  \SMALL $3\frac{1050094}{2639559}$  
    &  \SMALL $9\frac{41747}{879853}$, $2\frac{116513}{879853}$ \\
\hline
 \end{tabular}
\end{center}
 \caption{Data on trefoils of lengths $18$, $20$ and $22$
in the BCC lattice.}
  \label{tabletrefoilsbcc}
\end{table}
%%%%%%%%%%%%%%%%%%%%%%%%%%%%%%%%%%%%%%%%%%%%%%%%%%%%%%%%%%%%%%%%%%%%

%%%%%%%%%%%%%%%%%%%%%%%%%%%%%%%%%%%%%%%%%%%%%%%%%%%%%%%%%%%%%%
\subsubsection{The Lattice Writhe and Curvature:}

The average writhe $\C{W}_K$, the average absolute writhe $|\C{W}|_K$
and the average curvature $\C{K}_K$ (in units of $2\pi$)
of minimal length BCC lattice polygons are displayed in table
\ref{tableBCC}.  The writhe $W_r(\omega)$ of a BCC lattice polygon
$\omega$ is a rational number (since $12W_r(\omega)$ is an integer)
as shown in equation \Ref{eqn21AA}.  Thus, the average writhe and
average absolute writhe of minimal length BCC lattice polygons
are listed as rational numbers in table \ref{tableBCC}.  These
results are exact in those cases where we succeeded in finding
all minimal length BCC polygons of a particular knot type $K$.

The lattice curvature of a given BCC lattice polygon is somewhat
more complicated.  Each BCC lattice polygon $\omega$ has 
curvature which may be expressed in the form $B\arccos(1/\sqrt{3})
+ 2\pi C$, where $B$ and $C$ are rational numbers.  Thus, 
the average curvature of minimal BCC lattice polygons of knot type 
$K$ is given by expressions similar to equation \Ref{eqn21B},
with $\C{B}_K$ and $\C{K}_K$ rational numbers.  In table
\ref{tableBCC} the values of $\C{B}_K$ and $\C{K}_K$ are
given for each knot type, as a pair of rational numbers.  For example,
the average curvature of the unknot is $-2\arccos(1/\sqrt{3})
+3\pi = 7.51414\ldots > 2\pi$.  This shows that some minimal
conformations of the unknot are not planar.

Similarly, the average curvature of minimal length polygons of
knot type $3_1$ is given by $11\frac{4}{33}\arccos(1/\sqrt{3})
+\frac{21}{11}\pi = 16.6218\ldots < 6 \pi$.  In other words,
the average curvature of minimal length BCC lattice trefoils is
less than $6\pi$, which is the minimal lattice curvature of
SC lattice trefoils. In fact, one may check that this average
curvature is less than $5\frac{1}{2}\pi$, which is the average
curvature for minimal length FCC lattice polygons.  In other words,
the embedding of lattice trefoils of minimal length in the BCC
has lower average curvature than either the average curvature
in the SC or FCC lattices.

Similar to the results in the SC and FCC lattices, the
average and average absolute writhes in table \ref{tableSC} are 
equal, except in the case of achiral knots.  If $K$ is 
an achiral knot type, then generally $\C{W}_K = 0$ while 
$|\C{W}|_K > 0$, similar to the results in the SC and FCC lattices. 
Observe that the average absolute writhe of $8_3$  is zero in the 
BCC lattice, as it was in the SC lattice (but it is positive in the 
FCC lattice). 

The average writhe at minimal length of $3_1^+$ is 
\hbox{\small $3\frac{40}{99} \pi$} in the BCC lattice, which is
slightly smaller than the result in the SC lattice 
(\hbox{\small $3\frac{735}{1664}$}).  However,
it is still larger than the result in the FCC lattice.  Increasing 
the value of $n$ from $18$ to $20$ and $22$ in the BCC lattice 
and measuring the average writhe of $3_1^+$ gives the results in 
table \ref{tabletrefoilsbcc}, which shows that the average writhe 
decreases slowly with $n$, in contrast with the trend observed
in the FCC lattice. However, the average  writhe remains, as in the 
SC lattice, quite insensitive to $n$.

Generally, the average and average absolute writhe increases
with crossing number in table \Ref{tableSC}.  However, in each
class of knot types of crossing number $C>3$ there are knot types
with small average absolute writhe (and thus with small average
writhe).  For example, amongst the class of knot types on eight
crossings, there are achiral knots with zero absolute writhe
($8_3$ and $8_{12}$), as well as chiral knot types with average 
absolute writhe small compared to the average absolute writhe of (say)
$8_2$.  The knot types $8_4$, $8_8$, $8_9$, $8_{13}$, $8_{17}$ 
and $8_{18}$, amongst knot types on eight crossings, also have 
average absolute writhe less than $2$, which is small when compared 
to other eight crossing knots such as $8_2$.

The average curvature of minimal length BCC lattice knots are given 
in terms of the rational numbers $\C{B}_K$ and $\C{K}_K$,
as explained above.   Both $\C{B}_K$ and $\C{K}_K$ tends to 
increase with $n_K$ in table \ref{tableBCC}.  The ratios
$[\C{B}_K/n_K,\C{K}_K/n_K]$ however, may decrease with 
increasing $n_K$ within families of knot types. For example, for 
$(N,2)$-torus knots, these ratios decrease with increasing $n_K$ 
as $\{[0.618,0.053],[0.745,0.033],[0.770,0.031],[0.824,0.045]\}$
to three digits accuracy along the sequence $\{3_1,5_1,7_1,9_1\}$.
Similar patterns can be  determined for other families of knot 
types. 

Finally, the average curvature of the trefoil in the BCC lattice is
$11\frac{4}{33}\arccos(1/\sqrt{3}) +\frac{21}{11}\pi = 16.6218\ldots$
and this is less than the lower bound $6\pi$ of the minimal 
curvature of a trefoil in the SC lattice \cite{JvRP99}.  The
minimal curvature of $9_{47}$ at minimal length in the SC lattice 
is $9\pi$, but in the BCC our data show that the minimal 
curvature is $28.255468\ldots < 9\pi$.  In other words, there
are minimal length conformations of the knot $9_{47}$ in the
BCC lattice with total curvature less than the minimal
curvature $9\pi$ of this knot in the SC lattice \cite{JvRP99}.

%%%%%%%%%%%%%%%%%%%%%%%%%%%%%%%%%%%%%%%%%%%%%%%%%%%%%%%%%%%%%%%%%
%%%%%%%%%%%%%%%%%%%%%%%%%%%%%%%%%%%%%%%%%%%%%%%%%%%%%%%%%%%%%%%%%
\section{Conclusions}

Data for compounded lattice knots were significantly harder to collect
than for the prime knot types.  Thus, we collected data in only the
SC lattice, and we considered our data less secure if compared
to the data on prime knot types listed in tables
\ref{tableSC}, \ref{tableFCC} and \ref{tableBCC}.  The data
for compounded SC lattice knots are presented in table \ref{tableSCC}. 
Included are the first few members of sequences $\LL (3_1^+)^N \RR$
and $\LL 4_1^N \RR$ and mixed compound knots up to  eight
crossings, with $(3_1^+)^2 \# (3_1^-)$ included.  We made an
attempt to find all minimal knots of type $(3_1^+)^2 \# (3_1^-)^2$
but ran out of computer resources when $7000000$ symmetry classes
were detected.

%%%%%%%%%%%%%%%%%%%%%%%%%%%%%%%%%%%%%%%%%%%%
\begin{table}[h!]
\begin{center}
 \begin{tabular}{||c||r|r|r|r||l|l|l||}
 \hline
  Knot & \multicolumn{7}{|c||}{Simple Cubic Lattice} \\
  \hline  
   & $n_K$ & $\C{P}_K$ & \multicolumn{2}{|c||}{$\C{S}_K$} & $\C{W}_K$ & $\C{|W|}_K$ & $\C{K}_K$ \\
   \hline
$3_1^+$ & 24 &\SMALL  3328  &\SMALL  142  &\small  $24^{137}8^5$    
    &  \SMALL $3\frac{735}{1664}$ &  \SMALL $3\frac{735}{1664}$ 
    &  \SMALL $3 \frac{801}{1664}$ \\
$\LB 3_1^+ \RB^2$ & 40 &\SMALL  30576  &\SMALL  1275  &\small  $24^{1273}12^2$    
    &  \SMALL $6\frac{45}{49}$ &  \SMALL $6\frac{45}{49}$ 
    &  \SMALL $5 \frac{278}{637}$ \\
$\LB 3_1^+ \RB^3$ & 56 &\SMALL  288816  &\SMALL  12034  &\small  $24^{12034}$    
    & \SMALL $10\frac{1711}{4376}$ & \SMALL $10\frac{1711}{4376}$ 
    & \SMALL $7 \frac{13603}{48136}$ \\
$\LB 3_1^+ \RB^4$ & 72 &\SMALL  5582160  &\SMALL  232606  &\small  $24^{232582}8^{24}$   
    & \SMALL $13\frac{799403}{930360}$ 
    & \SMALL $13\frac{799503}{930360}$ 
    & \SMALL $9 \frac{410101}{930360}$ \\
$\LB 3_1^+ \RB^5$ & 88 &\SMALL  71561664  &\SMALL  2981736  &\small  $24^{2981736}$    
    & \SMALL $17\frac{4051667}{11926944}$ 
    & \SMALL $17\frac{4051667}{11926944}$ 
    & \SMALL $11 \frac{2709893}{3975648}$ \\
\hline
\SMALL $3_1^+\# 3_1^-$ & 40 &\SMALL  143904  &\SMALL  6058 &\small $24^{5934}12^{124}$    
    & \SMALL $0$ 
    & \SMALL $\frac{1085}{5996}$ 
    &  \SMALL $5 \frac{1749}{11992}$ \\
\SMALL $4_1 \# 3_1$ & 46 &\SMALL  359712 &\SMALL  14988 &\small $24^{14988}$    
    &  \SMALL $3\frac{4259}{9992}$ 
    &  \SMALL $3\frac{4259}{9992}$ 
    &  \SMALL $5 \frac{22693}{29976}$ \\
\SMALL $5_1^+\# 3_1^+$ & 50 &\SMALL  200976 &\SMALL  8374 &\small $24^{8374}$
    &  \SMALL $9\frac{10169}{16748}$ 
    &  \SMALL $9\frac{10169}{16748}$
    &  \SMALL $6\frac{10715}{16748}$ \\
\SMALL $5_1^+\# 3_1^-$ & 50 &\SMALL  568752 &\SMALL  23698 &\small $24^{23698}$
    &  \SMALL $2\frac{9174}{11849}$ 
    &  \SMALL $2\frac{9174}{11849}$
    &  \SMALL $6\frac{12351}{23698}$ \\
\SMALL $5_2^+\# 3_1^+$ & 52 &\SMALL  7357008 &\SMALL  306542 &\small $24^{306542}$
    &  \SMALL $7\frac{305199}{306542}$ 
    &  \SMALL $7\frac{305199}{306542}$
    &  \SMALL $6\frac{246067}{306542}$ \\
\SMALL $5_2^+\# 3_1^-$ & 50 &\SMALL  5280 &\SMALL  220 &\small $24^{220}$
    &  \SMALL $1\frac{7}{88}$ 
    &  \SMALL $1\frac{7}{88}$
    &  \SMALL $6\frac{87}{440}$ \\
\SMALL $(3_1^+)^2\# 3_1^-$ & 56 &\SMALL  8893152 &\SMALL  370548 &\small $24^{370548}$    
    &  \SMALL $3\frac{57571}{123516}$ 
    &  \SMALL $3\frac{57571}{123516}$ 
    &  \SMALL $6 \frac{167021}{185274}$ \\
\hline
$4_1$ & 30 &\SMALL  3648  &\SMALL  152 &\small $24^{152}$ 
    &  \SMALL 0 
    & \SMALL $\frac{33}{152}$ 
    &  \SMALL $4 \frac{1}{152}$ \\
$(4_1)^2$ & 52 &\SMALL  334824  &\SMALL  14144  &\small $24^{13758}12^{386}$    
    & \SMALL 0 
    & \SMALL $\frac{3450}{13951}$ 
    &  \SMALL $6 \frac{13089}{27902}$ \\
$(4_1)^3$ & 74 &\SMALL  31415592 &\SMALL  1308983  &\small $24^{1308983}$    
    & \SMALL 0 
    & \SMALL $\frac{3384343}{1308983}$ 
    &  \SMALL $8 \frac{2294775}{2617966}$ \\
\hline
 \end{tabular}
\end{center}
 \caption{Data on knots in the SC Lattice.}
  \label{tableSCC}
\end{table}
%%%%%%%%%%%%%%%%%%%%%%%%%%%%%%%%%%%%%%%%%%%%%%%%%%%%%%%%%

Compound knot types in the SC lattice tended to have
far larger numbers of symmetry classes at minimal length,
compared to prime knot types with similar minimal length or
crossing numbers.  We ran our simulations for up to weeks
in some cases, in an attempt to determine good bounds
on the numbers of minimal length polygons. As in the case
of prime knots, certainty about our data decreases with 
increasing numbers of symmetry classes, from very certain when
$\C{P}_K \lesssim 1000$, to reasonably certain when the
number exceeds $1000 \lesssim \C{P}_K \lesssim 10000$, 
less certain when $10000 \lesssim \C{P}_K \lesssim 100000$, and
the stated value of $\C{P}_K$ should be considered a lower
bound if $\C{P}_K \gtrsim 100000$. 

That is, the data in table \ref{tableSCC} for the knot 
types $\LL (3_1^+)^4 \RR$, $\LL (3_1^+)^5 \RR$ and $\LL 4_1^3 \RR$
may not be exact for $\C{P}_K$, symmetry classes and estimates 
of the writhe and curvature.  At best, those results are lower 
bounds on the counts, within a few percent of the true results.

The data in table \ref{tableSCC} allows us to make rough
estimates of $\gamma_{3_1}$ (see equation
\Ref{eqn14}). By taking logarithms of $\C{P}_{3_1}$, one gets 
for $\gamma_{3_1}$ the following estimates with increasing $N$: 
$\{8.1101,5.1639,4.1912,3.8838,3.4424\}$.  These values have
not settled down and it is apparent that simulations with more
complex compounded knots will be needed to estimate $\gamma_K$.

In addition, we can make estimates analogous to $\gamma_K$ 
by considering the writhe or curvature instead:  Define
\begin{equation}
\zeta_K = \limsup_{N\to\infty} \frac{\LV \C{W} \RV_{K^N}}{N}
\quad\hbox{and}\quad
\beta_K = \limsup_{N\to\infty} \frac{\C{K}_{K^N}}{N},
\end{equation}
then one may attempt to estimate these numbers for the
trefoil and figure eight knots.  $\zeta_K$ can
be interpreted as the average absolute writhe per knot component, 
and similarly, $\beta_K$ is the average curvature at minimal 
length per knot component.

The data for the trefoil give the sequence
$\{3.441,3.459,3.464,3.465,3.468\}$.  These results show that
$\zeta_{3_1} \approx 3.47$.  The similar analysis for
$4_1$ gives $\{0.217,0.124,0.099\}$ and it appears that there is a
more pronounced dependence on the number of components in this case.
It is difficult to estimate $\zeta_{4_1}$ from these results,
and we have not ruled out the possibility that it may approach
zero as the number of components increases without bound.

%%%%%%%%%%%%%%%%%%%%%%%%%%%%%%%%%%%%%%%%%%%%
\begin{table}[h!]
\begin{center}
 \begin{tabular}{||c||c|c|c||}
 \hline
Rank & SC Lattice & FCC Lattice & BCC Lattice \\
  \hline  
1 & \small $3_1^+$ & \small $0_1$ & \small $0_1$ \\
2 & \small $5_2^+$ & \small $8_{20}^+$ & \small $3_1^+$ \\
3 & \small $0_1$ & \small $6_3$ & \small $5_1^+$ \\
4 & \small $4_1$ & \small $4_1$ & \small $8_{20}^+$ \\
5 & \small $7_2^+$ & \small $8_1^+$ & \small $5_2^+$ \\
6 & \small $6_2^+$ & \small $8_{11}^+$ & \small $8_2^+$ \\
7 & \small $5_1^+$ & \small $6_1^+$ & \small $8_4^+$ \\
8 & \small $8_{21}^+$ & \small $8_{13}^+$ & \small $6_2^+$ \\
9 & \small $7_1^+$ & \small $8_2^+$ & \small $7_3^+$ \\
10 & \small $8_2^+$ & \small $8_{21}^+$ & \small $6_3$ \\
11 & \small $7_6^+$ & \small $7_5^+$ & \small $8_7^+$ \\
12 & \small $8_{19}^+$ & \small $8_{18}$ & \small $8_8^+$ \\
13 & \small $8_9$ & \small $8_4^+$ & \small $7_4^+$ \\
14 & \small $6_3$ & \small $8_3$ & \small $7_5^+$ \\
15 & \small $8_{15}^+$ & \small $6_2^+$ & \small $8_{13}^+$ \\
16 & \small $8_4^+$ & \small $8_{15}^+$ & \small $8_{14}^+$ \\
17 & \small $6_1^+$ & \small $8_{12}$ & \small $8_{12}^+$ \\
18 & \small $8_{17}^+$ & \small $8_{14}^+$ & \small $8_6^+$ \\
19 & \small $8_{13}^+$ & \small $7_1^+$ & \small $8_{18}$ \\
20 & \small $8_1^+$ & \small $5_2^+$ & \small $8_{10}^+$ \\
21 & \small $7_5^+$ & \small $8_{16}^+$ & \small $8_9^+$ \\
22 & \small $8_6^+$ & \small $7_2^+$ & \small $8_5^+$ \\
23 & \small $7_4^+$ & \small $7_6^+$ & \small $7_1^+$ \\
24 & \small $8_{18}$ & \small $8_9^+$ & \small $8_{15}^+$ \\
25 & \small $8_8^+$ & \small $8_6^+$ & \small $8_{19}^+$ \\
26 & \small $8_{12}$ & \small $3_1^+$ & \small $8_{17}^+$ \\
27 & \small $8_{10}^+$ & \small $8_{17}^+$ & \small $8_{21}^+$ \\
28 & \small $8_5^+$ & \small $8_5^+$ & \small $8_{16}^+$ \\
29 & \small $7_7^+$ & \small $8_8^+$ & \small $8_1^+$ \\
30 & \small $7_3^+$ & \small $7_3^+ $ & \small $8_3$ \\
31 & \small $8_{14}^+$ & \small $7_7^+$ & \small $6_1^+$ \\
32 & \small $8_{20}^+$ & \small $8_{10}^+$ & \small $4_1$ \\
33 & \small $8_{11}^+$ & \small $8_7^+$ & \small $7_6^+$ \\
34 & \small $8_{16}^+$ & \small $8_{19}^+ $ & \small $7_2^+,\,7_7^+$ \\
35 & \small $8_7^+$ & \small $5_1^+$ & \small $8_{12}$ \\
36 & \small $8_3$ & \small $7_4^+$ & \small $ $ \\
\hline
 \end{tabular}
\end{center}
 \caption{Ranking of minimal length lattice knots of knot 
types to 8 crossings by entropy per unit length in the SC, 
FCC and FCC lattices.}
  \label{tableME}
\end{table}
%%%%%%%%%%%%%%%%%%%%%%%%%%%%%%%%%%%%%%%%%%%%%%%%%%%%%%%%%

Repeating the above for $\beta_{3_1}$ gives the estimates
$\{3.481,2.718,2.428,2.360,2.336\}$ so that one may estimate
$\beta_{3_1} \approx 2.3$ in units of $2\pi$. 
Observe that the bounds on $\nu_{3_1}$
following equation \Ref{eqn25} suggest that $\beta_K \geq 1$
for any knot type $K \not= 0_1$. The estimates for $4_1$ are
$\{4.007,3.336,2.960\}$, so that one cannot yet determine 
an estimate for $\beta_{4_1}$.

Overall we have examined the entropic and average geometric
properties of minimal length lattice knots in the SC, the FCC
and the BCC lattices.  Our data were collected using Monte Carlo
algorithms with BFACF-style elementary moves. The statistical
and average properties of sets of minimal length knotted polygons
were determined and discussed, and comparisons were made between
the results in the three lattices.  Our results show in particular
that the properties of minimal length lattice knots are not
universal in the three lattices.  The spectrum of minimal length
knot types, the entropy, and the average lattice curvature and  
lattice writhe shows variation in several aspects.  For example,
the spectra of minimal length knots in tables \ref{tablenKSC},
\ref{tablenKFCC} and \ref{tablenKBCC} do not maintain a strict
order, but shuffle some knots types up or down the table in
the different lattices.

Similar observations can be made with respect to the entropy
of minimal length knots.  For example the entropy of 
the knot types $5_1$ and $5_2$ are inverted in the BCC lattice,
compared to the relation they have in the SC and FCC lattices. 
In table \ref{tableME} we rank knot types by the entropy per unit length
at minimal length.  That is, we rank the knot types by computing
$\C{E}_K$ (see equation \Ref{eqn31}) -- the larger the result,
the lower the ranking in the table (that is, the higher the knot type
is listed in the table).  The rankings in table \ref{tableME} 
are shuffled around in each of the three lattices.  For example,
the trefoil knot is ranked at position $1$ in the SC lattice, at 
position $26$ in the FCC lattice, and at position $2$ in the BCC lattice.
Other knot types are similarly shuffled.

In the case of writhe there are also subtle, but interesting 
differences between the three lattices.  For example, the
average absolute writhe of the knot type $8_3$ is zero in the
SC and BCC lattices, yet it is not zero in the FCC lattice. 
Equally interesting about the results in the FCC lattice is
the fact that the average absolute writhes of the knot types $4_1$ 
and $8_{16}$ are very nearly very simple fractions (far simpler than
the number of symmetry classes in each case would suggest), in
addition to the fact that the average absolute writhe of the
knot $4_1$ is identically zero in the BCC lattice (but not
in the SC and FCC lattices).

Finally, an analysis of the number of knot types of minimal
length $n_K\leq n$, denoted $Q_n$, suggest that $Q_n \sim
\C{Q}^n$.  Our data suggest that $\C{Q}_{SC} < \C{Q}_{BCC}
< \C{Q}_{FCC}$, so that that the number of knot types which
can be tied in a polygon of $n$ edges increases fastest
(at an exponential rate) in the FCC lattice, followed by the 
BCC lattice and then the SC lattice.

%%%%%%%%%%%%%%%%%%%%%%%%%%%%%%%%%%%%%%%%%%%%%%%%%%%%%%%%%%%%%%%%
%%%%%%%%%%%%%%%%%%%%%%%%%%%%%%%%%%%%%%%%%%%%%%%%%%%%%%%%%%%%%%%%

\section*{Acknowledgments}
The authors acknowledge funding in the from of Discovery
Grants from NSERC (Canada).

\vspace{5mm}

\vspace*{0.5cm}

\noindent{\bf{Bibliography}}

\vspace*{0.3cm}

\begin{center}

\end{center}

% Set the ending of a LaTeX document
\end{document}